\documentclass[numberedappendix]{emulateapj}

\usepackage{amsmath}
\usepackage{color}
\usepackage{float}

\def\be{\begin{equation}}
\def\ee{\end{equation}}
\def\ba{\begin{eqnarray}}
\def\ea{\end{eqnarry}}
\def\bal#1\eal{\begin{align}#1\end{align}}

\shorttitle{A line-smeared treatment of opacities}

\begin{document}
\title{A line-smeared treatment of opacities for the spectra and\\
light curves from macronovae}

\author{Christopher J. Fontes$^{1}$, Chris L. Fryer$^{1,2,3}$,
Aimee L. Hungerford$^{1}$, Ryan T. Wollaeger$^{1}$,\\
Stephan Rosswog$^{4}$, and Edo Berger$^{5}$}
\affil{$^1$Los Alamos National Laboratory, Los Alamos, NM 87545, USA\\
$^2$Physics Department, University of Arizona, Tucson, AZ 85721, USA\\
$^3$Physics and Astronomy Department, University of New Mexico, Albuquerque,
NM 87131, USA\\
$^4$Astronomy and Oskar Klein Centre, Stockholm University, AlbaNova,
SE-10691 Stockholm, Sweden\\
$^5$Harvard-Smithsonian Center for Astrophysics,
60 Garden Street, Cambridge, MA 02138, USA}


\begin{abstract}

Gravitational wave observations need accompanying electromagnetic
signals to accurately determine the sky positions of the sources. The
ejecta of neutron star mergers are expected to produce such
electromagnetic transients, called macronovae.  Characteristics of the
ejecta include large velocity gradients and the presence of heavy
$r$-process elements, which pose significant challenges to the
accurate calculation of radiative opacities and radiation
transport. For example, these opacities include a dense forest of
bound-bound features arising from near-neutral lanthanide and actinide
elements. Here we investigate the use of fine-structure,
line-smeared opacities that preserve the integral of the opacity
over frequency, which is an alternative approach to the expansion-opacity
formalism. The use of area-preserving line profiles
produces frequency-dependent opacities that are one to
two orders of magnitude greater than those obtained with the use of
expansion opacities in the Sobolev approximation.  Opacities are
calculated for four $r$-process elements (cerium, neodymium, samarium
and uranium) using fully and semi-relativistic methods, as well as
different amounts of configuration interaction in the atomic
structure, in order to test the sensitivity of the emission to the
underlying atomic physics.  We determine the effect of these new
opacities on simulated spectra and broad-band light curves by applying
a multi-dimensional ray-trace method to the ejecta predicted from
3-dimensional merger calculations.  With our substantially larger
opacities the simulations yield slightly lower luminosities that peak
in the mid-IR. These results suggest that those radioactively powered
transients that are related to the very heaviest $r$-process material
are more difficult to observe than previously believed.

\end{abstract}



\section{Introduction}
\label{sec:intro}

With the recent announcement of the first direct observation of
gravitational waves (GW; \citealt{gwave1,gwave3,gwave2}) the era of
gravitational wave astronomy has begun. This discovery provides urgent
motivation for a deeper understanding of the physics of the GW sources
and their evolutionary paths. While the observed merger of two
black holes is from a gravitational point of view the cleanest
possible system, the lack of an accompanying electromagnetic (EM)
signal\footnote{But see the recent controversy
  \citep{connaughton16,lyutikov16,greiner16} about a claimed EM counterpart.}
leaves us blind with respect to the astrophysical environment of the
source. Although most detected GW sources are nowadays believed to be black
hole binaries \citep{belczynski10,dominik15}, neutron star mergers
(NSMs) are another major source of gravitational waves for which an EM signal
is clearly expected. The short-term EM waves are predicted to occur in two
forms: gamma-ray bursts (GRBs) \citep{eichler89,kouveliotou93,berger14}
and dim, supernova-like events
\citep{li_nsm98}, sometimes referred to as macronovae
\citep{kulkarni_nsm05} or kilonovae \citep{metzger10}. 
The GRBs are expected to be short-lived with a beamed signal, offering a
relatively low probability of being observed (see, for example,
\citealt{fong12,berger13,fong14,grossman14,fong15}). On the other hand,
macronovae\footnote{In this work,
we favor the more generic term ``macronova" over ``kilonova" because simulated
luminosities have moved progressively lower than the original estimate
of 1,000 times that associated with a standard nova.},
which are the focus of the present study, are predicted to generate
relatively isotropic signals and therefore offer a major advantage for
coincident observation with gravitational waves.

An understanding of the EM emission produced by macronovae requires detailed
information and simulations, such as the composition, velocity
structure and opacity of
the ejecta, and the transport of radiation through this material.
Theoretical investigations of these transients have undergone steady
improvement in various respects, especially the characterization
of the opacity, which we summarize here.
Early light-curve estimates \citep{li_nsm98} were determined from a constant
opacity, 0.2~cm$^2$/g, due to electron scattering.
That well-known gray opacity value results from the Thomson-scattering opacity,
$\kappa^{\rm T} = 0.4 (\overline{Z}/A)$~(cm$^2$/g),
under the assumption of fully ionized material, i.e. $\overline{Z} = Z$.
Here, $Z$ is the atomic number, $A$ the atomic weight,
$\overline{Z}$ the mean ion charge, and $Z/A\approx 1/2$ for most elements.
An improvement over this approach was to approximate the complex opacity
of the $r$-process elements in the ejecta using bound-bound lines from
iron, along with corrections to the ionization energies \citep{metzger10}.

Significantly improved opacities were subsequently obtained by the Mons group
\citep{biemont99,quinet99,palmeri00,biemont04,quinet04}
using modern atomic structure calculations of neutral and near-neutral
lanthanide species (i.e. the D.R.E.A.M. database). Their approach employs the
least-squares fitting procedure in the Cowan code \citep{cowan} in conjunction
with a core-polarization potential to obtain spectroscopic quality radiative
rates. However, this approach requires access to accurate level energies,
which are not readily available for the large number of levels required
to construct complete atomic models for macronova simulations.
A more recent approach used ab initio atomic structure calculations
of $r$-process elements in combination with the expansion-opacity
approach \citep{sobolev60,castor74,karp77} to produce detailed
spectra and light curves \citep{kasen13}. The latter investigation included
neodymium as a representative ejecta element, using realistic
near-neutral ion species, rather than the assumption of full
ionization. A complementary study was performed using opacities
composed of all the $r$-process elements \citep{tanaka13}.
This progression in opacity improvements moved
the prediction of peak emission to progressively longer wavelengths, 
with current best estimates going as low as the near-infrared band
\citep{barnes13}.
Recent modeling has taken into account these more realistic
opacities to study various aspects of macronovae. For example,
a more sophisticated treatment of the ejecta geometry has been considered
in 3D hydrodynamic simulations \citep{rosswog14,grossman14}.
This study used an enhanced gray opacity for the ejecta, 10~cm$^2$/g,
and explored in a simple model the possibility of a second transient
due to neutrino-driven winds
with an opacity of 1~cm$^2$/g to represent elements of intermediate
mass between iron and the lanthanides.
Such wind transients have been explored in more detail recently based
on 3D hydrodynamical simulations \citep{perego14,martin15}.

The opacity of $r$-process ejecta in NSMs, however,
remains a significant source of uncertainty in the simulation of
macronova emission.  A single ion stage for a lanthanide or actinide
element can produce in excess of $\sim 10^8$ lines for the conditions
of interest (see Section~\ref{sec:atomic}).  While it is impossible to
calculate all such lines with spectroscopic accuracy, the presence of large
velocity gradients might allow simplifications in the opacity
calculations.  The expansion-opacity formalism developed to study
Type~Ia
supernovae~\citep[e.g.,]{eastpinto93,fryer99,pintoeast00,kasen06}
is based on the argument that the opacity can be modeled by a finite number
of distinct
strong lines.  If the photon is re-emitted at the same energy as it is
absorbed, we can treat each line as a scattering center.  The strong
velocity gradients mean that, as this photon travels, the Doppler
effect on its energy with respect to the rest frame of the matter
shifts it off the line center, allowing it to once again transport until it
is Doppler boosted into another line
\citep{sobolev60,castor74,karp77}.
Among the simplifying assumptions
in this method are that the lines dominating the opacity are
sufficiently distinct, such that radiation transport for
each line can be treated individually,
and there is no line splitting, also known as fluorescence
(i.e. no altering of the photon energy
due to emission via a transition different from the absorbing one).  In
general, these assumptions do not hold for most supernova
calculations.  Where they are valid, perhaps for some wavelengths and
times in type~Ia supernovae~\citep{pintoeast00}, this expansion
opacity greatly simplifies the opacity calculations.

\begin{figure}[h]
\centering
\includegraphics[clip=true,angle=0,width=0.8\columnwidth]{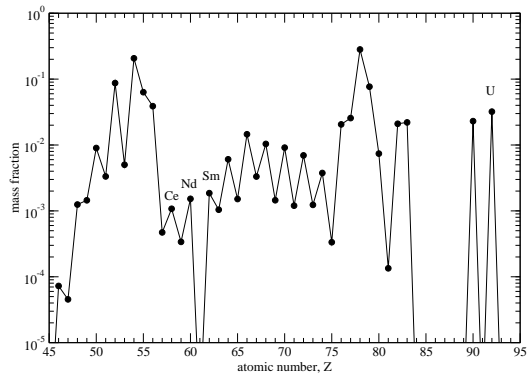}
\caption{Predicted elemental abundances for the ejecta produced
in a $1.4+1.4\,{\rm M_\odot}$ neutron star merger~\citep{rosswog14,grossman14}.
Labels are provided for the four elements considered in this work.
}
\label{fig:mfrac}
\end{figure}

For the dense forest and complex structure of lines for the $r$-process
elements, this expansion approximation may not be valid.  Not only do
these elements produce a dense cluster of lines in the optical and near
infra-red bands (a single lanthanide or actinide element can produce one to
two orders of magnitude more lines than iron or silicon), but the opacity is
dominated by a broad set of elements, rather than a few.
For example, in our NSMs, a significant number of lanthanide and actinide
elements are predicted to be present in the ejecta (see
Figure~\ref{fig:mfrac}), which are expected to provide the dominant
contribution to the opacity.

In earlier works, the process of fluorescence was found to be high
and was included to modify
to the expansion opacities in order to obtain improved agreement with
observations for Type~Ia supernovae~\citep{pintoeast00,kasen06}.
Here, we investigate the possibility that fluorescence
is sufficiently strong in the NSM ejecta to destroy the coherent scattering
associated with multiple photon scatterings within a line.
We explore such an approach by smearing the lines in
a manner that conserves the total integral of the frequency-dependent
opacity~\citep{fontes15}. For computational convenience we perform the smearing
via a Doppler width in the Voigt line profile, choosing a characteristic value
that is consistent
with the spatial resolution employed in our light-curve simulations.
Of course, Doppler broadening is formally associated with absorbers undergoing
random motion, as opposed to the ordered, homologous expansion of the ejecta.
Also, the Voigt profile is symmetric about the line center, while the
opacity-expansion formalism combined with the
homologous flow assumes that the photons undergo only redward shifts.
Nevertheless, we expect our approach to take into account, in an
average sense, the fact that a radial zone, which is characterized by a
zone-centered velocity, actually encompasses a range of velocities, with
photons that are both redder and bluer relative the their energy
at the zone center (see Section~\ref{subsec:comp_spec}).

This line-smearing approach requires detailed opacity 
calculations for all lines and is computationally intensive,
but it automatically takes into account the case of strong line splitting.
Detailed atomic physics models are presented for three lanthanides
and one actinide in Section~\ref{sec:atomic}:
Ce $(Z=58)$, Nd $(Z=60)$, Sm $(Z=62)$ and U $(Z=92)$.
We provide several different atomic physics models to assess the
sensitivity of the opacities, light curves and spectra to the quality of the
atomic data. Line-smeared opacities are presented for these four elements
in Section~\ref{sec:opac}. As expected, the resulting opacities are
significantly enhanced (by one to two orders of magnitude) compared to those
obtained via the Sobolev reduction employed in the opacity-expansion formalism.

In Section~\ref{sec:lc} we use a 2-dimensional ray-trace transport
code to determine the effect of different variants of the opacity
calculations on the spectra and broad-band light curves from the
ejecta of three merger models~\citep{rosswog14}, mapping these 3-dimensional
models assuming cylindrical symmetry.  We summarize with a brief 
discussion of the implications of our results for observations.

\section{Atomic Physics Considerations}
\label{sec:atomic}

\subsection{Computational Framework}
\label{subsec:compframe}

In this work we use the Los Alamos suite of atomic physics and plasma modeling
codes~(see \citealt{LANL_suite} and references therein)
to generate the fundamental data and opacities needed
to simulate the characteristics (time to peak, spectra, luminosities,
decay times, etc.) associated with neutron star mergers.
For a given element, a model is composed of the atomic structure
(energies, wavefunctions and oscillator strengths) and photoionization
cross sections. Both the fully and semi-relativistic capabilities of the suite
are used in this work.

The fully relativistic (FR) approach is based on bound- and continuum-electron
wavefunctions that are solutions of the Dirac equation, while the
semi-relativistic (SR) approach uses solutions of the 
Schr\"odinger equation with relativistic corrections.
A fully relativistic (FR) calculation begins with the RATS atomic
structure code \citep{LANL_suite} using the Dirac-Fock-Slater
method of Sampson and co-workers \citep{sampson_physrep}.
A semi-relativistic (SR) calculation begins with the CATS atomic
structure code \citep{cats_man} using the Hartree-Fock method
of Cowan~\citep{cowan}.
These calculations produce detailed, fine-structure data that include a
complete description of configuration interaction for the specified list of
configurations.
Two variant, relativistic calculations are also considered that include
incomplete amounts of configuration interaction
(see Section~\ref{subsec:models_base}).
After the atomic structure calculations are complete,
both the FR and SR methods use the GIPPER ionization code
to obtain the relevant photoionization cross sections
in the distorted-wave approximation. The photoionization data are used
to generate the bound-free contribution to the opacity and are not
expected to be too important for the present application, due to
the range of relevant photon energies, but are included
for completeness. Therefore, they are calculated in the configuration-average
approximation, rather than fine-structure detail, in order to minimize the
computational time.

The atomic level populations are calculated with the ATOMIC code
from the fundamental atomic data. The code can be used in either
local thermodynamic equilibrium (LTE) or non-LTE mode
\citep{atomic1,hakel06,colgan_oplib,fontes_cr16}. The LTE approach was chosen
for the present
application, which requires only the atomic structure data in calculating
the populations. At the relevant times, the ejecta densities are low enough
that collective, or plasma, effects are not important and simple
Saha-Boltzmann statistics is sufficient to produce accurate level
populations. The populations are then combined with the oscillator
strengths and photoionization cross sections in ATOMIC to obtain
the opacities, which are constructed from the standard four contributions:
bound-bound (b-b), bound-free (b-f), free-free (f-f) and scattering.
Specific formulae for these contributions are readily available in various
textbooks, such as \citet{huebner}.
Here, we reproduce only the expression for the bound-bound
contribution, as it is useful for the subsequent discussion
of line-smeared opacities in Section~\ref{sec:opac}:
\begin{equation}
\kappa^{\rm b-b}_\nu = \frac{\pi e^2}{\rho m_e c} \sum_i N_i
\, |f_{i}| \, L_{i,\nu} \,,
\label{opac_bb}
\end{equation}
where $\nu$ is the photon energy,
$\rho$ is the mass density, $N_i$ is the number density of the initial
level in transition $i$,
$f_{i}$ is the oscillator strength describing the photo-excitation
of transition $i$, and $L_{i,\nu}$ is the corresponding line profile function.

\subsection{Baseline Atomic Models}
\label{subsec:models_base}

As mentioned in Section~\ref{sec:intro}, atomic models were created for four
elements: Ce $(Z=58)$, Nd $(Z=60)$, Sm $(Z=62)$, and U $(Z=92)$.
In order to obtain converged opacities for the range of
temperatures and densities in our simulations, only the first four
ion stages of each element were considered, similar to the choice
made by \citet{kasen13}. A list of configurations chosen for each element
is provided in Table~\ref{tab:configs}.
\begin{table*}
\centering
\caption{\rm A list of configurations, number of fine-structure levels,
and number of (electric dipole) absorption lines for the various ion stages
considered in this work. A completely filled Xe core is assumed
for Ce, Nd and Sm, while a filled Rn core is assumed for U.}
\vspace*{0.5\baselineskip}
\begin{tabular}{lcrr}
\hline
Ion stage   &  Configurations &  \# of levels & \# of lines \\
\hline
Ce {\sc i} &
$4f^2 6s^2$
$4f^1 5d^1 6s^2$,
$4f^1 5d^2 6s^1$,
$4f^1 5d^1 6s^1 6p^1$,
$4f^2 5d^1 6s^1$, & 2,546 & 626,112 \\
&
$4f^2 6s^1 6p^1$,
$4f^1 5d^3$,
$4f^1 6s^2 6p^1$,
$4f^1 5d^2 6p^1$,
$4f^2 5d^2$,
$4f^2 5d^1 6p^1$ \\

Ce {\sc ii} &
$4f^2 6s^1$,
$4f^2 5d^1$,
$4f^2 6p^1$,
$4f^1 5d^2$,
$4f^1 6s^2$, & 519 & 28,887 \\
&
$4f^1 5d^1 6s^1$,
$4f^1 5d^1 6p^1$,
$4f^1 6s^1 6p^1$,
$5d^3$,
$4f^3$ \\

Ce {\sc iii} &
$4f^2 $,
$4f^1 6s^1$,
$4f^1 5d^1$,
$4f^1 6p^1$,
$5d^2$,
$5d^1 6s^1$ & 62 & 452 \\

Ce {\sc iv} &
$4f^1$,
$6s^1$,
$5d^1$,
$6p^1$ & 7 & 8\\
\hline

Nd {\sc i} &
$4f^4 6s^2$,
$4f^3 5d^1 6s^2$,
$4f^4 5d^1 6s^1$,
$4f^4 5d^2$, & 18,104 & 25,224,451 \\
&
$4f^3 5d^1 6s^1 6p^1$,
$4f^4 5d^1 6p^1$,
$4f^4 6s^1 6p^1$ \\

Nd {\sc ii} &
$4f^4 6s^1$,
$4f^4 5d^1$,
$4f^4 6p^1$,
$4f^3 5d^2$, & 6,888 & 3,958,977 \\
&
$4f^3 5d^1 6s^1$,
$4f^3 5d^1 6p^1$,
$4f^3 6s^1 6p^1$ \\

Nd {\sc iii} &
$4f^4$,
$4f^3 6s^1$,
$4f^3 5d^1$,
$4f^3 6p^1$, & 1,650 & 233,822 \\
&
$4f^2 5d^2$,
$4f^2 5d^1 6s^1$,
$4f^1 5d^2 6s^1$ \\

Nd {\sc iv} &
$4f^3 $,
$4f^2 6s^1$,
$4f^2 5d^1$,
$4f^2 6p^1$, & 241 & 5,784 \\
\hline

Sm {\sc i} &
$4f^6 6s^2$,
$4f^5 5d^1 6s^2$,
$4f^6 5d^1 6s^1$,
$4f^6 5d^2$, & 60,806 & 249,301,825 \\
&
$4f^5 5d^1 6s^1 6p^1$,
$4f^6 5d^1 6p^1$,
$4f^6 6s^1 6p^1$\\

Sm {\sc ii} &
$4f^6 6s^1$,
$4f^6 5d^1$,
$4f^6 6p^1$,
$4f^5 5d^2$, & 29,970 & 67,743,385 \\
&
$4f^5 5d^1 6s^1$,
$4f^5 5d^1 6p^1$,
$4f^5 6s^1 6p^1$ \\

Sm {\sc iii} &
$4f^6$,
$4f^5 6s^1$,
$4f^5 5d^1$,
$4f^5 6p^1$, & 13,170 & 13,318,114 \\
&
$4f^4 5d^2$,
$4f^4 5d^1 6s^1$,
$4f^3 5d^2 6s^1$ \\

Sm {\sc iv} &
$4f^5$,
$4f^4 6s^1$,
$4f^4 5d^1$,
$4f^4 6p^1$ & 1,994 & 320,633 \\
\hline

U {\sc i} &
$5f^4 7s^2$,
$5f^3 6d^1 7s^2$,
$5f^4 6d^1 7s^1$, & 16,882 & 20,948,831 \\
&
$5f^4 6d^2$,
$5f^3 6d^1 7s^1 7p^1$,
$5f^4 6d^1 7p^1$\\

U {\sc ii} &
$5f^3 7s^2$,
$5f^4 7s^1$,
$5f^4 6d^1$,
$5f^4 7p^1$, & 6,929 & 4,016,742 \\
&
$5f^3 6d^2$,
$5f^3 6d^1 7s^1$,
$5f^3 6d^1 7p^1$,
$5f^3 7s^1 7p^1$\\

U {\sc iii} &
$5f^4$,
$5f^3 7s^1$,
$5f^3 6d^1$,
$5f^3 7p^1$, & 1,650 & 233,822 \\
&
$5f^2 6d^2$,
$5f^2 6d^1 7s^1$,
$5f^1 6d^2 7s^1$\\

U {\sc iv} &
$5f^4 7s^2$,
$5f^3 6d^1 7s^2$,
$5f^4 6d^1 7s^1$, & 241 & 5,784 \\
&
$5f^4 6d^2$,
$5f^3 6d^1 7s^1 7p^1$,
$5f^4 6d^1 7p^1$\\
\hline
\end{tabular}
\label{tab:configs}
\end{table*}
Since Nd was used as a representative element
in the recent study by \citet{kasen13}, we chose an identical
list of configurations for that element in order to make meaningful
comparisons. As expected, the number of Nd levels is identical to those
appearing in Table~1 of that earlier work.\footnote{Based on this analysis,
the $4f^4 6s^1 6p^1$ configuration appears to have been left out of
Table~1 of \citet{kasen13}.}
The number of lines is slightly higher in the present listing, possibly due
to the retention of small oscillator strengths that do not affect
the modeling in a significant way.
We did some tests to include higher lying configurations, but found that
the displayed list is sufficient to produce converged opacities due to
the relatively low temperature and densities of the ejecta.
Therefore, the configuration lists for the other elements were chosen
in a similar fashion. The configurations for Ce~{\sc ii} and Ce~{\sc iii}
are also identical to those chosen by \citet{kasen13}. The number of
levels and lines differ strongly in this case. We cross-checked
the values between our FR and SR calculations, which agree well,
so those earlier values appear to be in error.

\begin{figure}
\centering
\includegraphics[clip=true,angle=0,width=0.9\columnwidth]{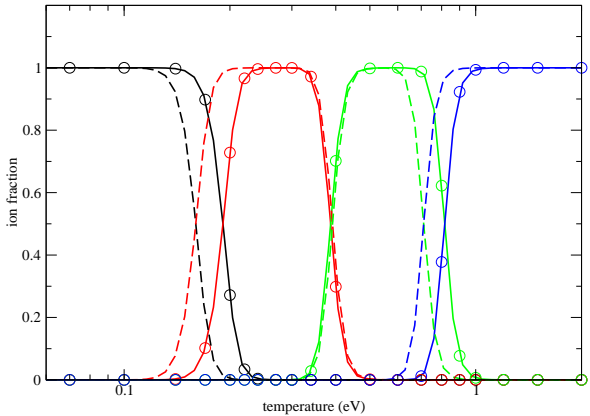}
\caption{Ionization-stage fraction versus temperature for Nd at a typical mass
density of $\rho = 10^{-13}$~g/cm$^3$, calculated with the fully relativistic
(FR) approach. The black curves and circles refer to Nd {\sc i}, the
red ones to Nd {\sc ii}, the green ones to Nd {\sc iii}, and
the blue ones to Nd {\sc iv}. The solid curves use NIST-corrected
ionization energies~\citep{nist}, while the dashed curves use
uncorrected values. The circles
indicate explicit temperatures at which (NIST-corrected) opacities were
calculated for use in the simulation of spectra and light curves.
}
\label{fig:ionfrac_nd}
\end{figure}

As a proxy for the quality of our atomic structure calculations,
the ionization energies for the FR and SR models are presented
in Table~\ref{tab:ionpot}, along with the values from the NIST
database \citep{nist}.
\begin{table}[h]
\centering
\caption{\rm Ionization energies for the first three
ion stages of each element considered in this work. Values are presented
for the fully relativistic (FR) and semi-relativistic (SR) methods described
in the text, as well as from the NIST database \citep{nist}.}
\vspace*{0.5\baselineskip}
\begin{tabular}{lccc}
\hline
Ion stage   &   \multicolumn{3}{c}{Ionization energy (eV)} \\
\cline{2-4}
            &  FR  &  SR  & NIST \\
\hline
Ce {\sc i}  & 4.91 & 5.24 & 5.54 \\
Ce {\sc ii} & 10.5 & 11.2 & 10.9 \\
Ce {\sc iii}& 18.0 & 19.6 & 20.2 \\
\hline
Nd {\sc i}  & 4.58 & 4.97 & 5.53 \\
Nd {\sc ii} & 10.9 & 11.1 & 10.7 \\
Nd {\sc iii}& 19.2 & 20.5 & 22.1 \\
\hline
Sm {\sc i}  & 4.83 & 5.33 & 5.64 \\
Sm {\sc ii} & 10.6 & 10.7 & 11.1 \\
Sm {\sc iii}& 20.3 & 21.6 & 23.4 \\
\hline
U {\sc i}  & 4.00 & 5.48 & 6.19 \\
U {\sc ii} & 12.1  & 11.7 & 11.6 \\
U {\sc iii}& 17.4 & 19.2 & 19.8 \\
\hline
\end{tabular}
\label{tab:ionpot}
\end{table}
The overall agreement is good, with the worst comparisons occurring
for the neutral ion stage (particularly for uranium), which is typically
the most difficult to calculate
due to the presence of more bound electrons and the need
to accurately describe the correlation between them.
As expected, the SR values are more accurate than the FR values
for these near-neutral ions for two reasons.
First, the Hartree-Fock approach uses a better (non-local) description
of the exchange interaction between the bound electrons than the 
Dirac-Fock-Slater method, which uses the Kohn-Sham local-exchange
approximation \citep{kohnsham,sampson_physrep}.
Second, the SR approach
uses semi-empirical scale factors to modify the radial integrals that
appear in the configuration-interaction calculation \citep{cowan}.
In any event, the inaccuracies
in the ionization energies have been removed in both the FR and SR
models of the present study by replacing the calculated values with
those appearing in the NIST database. All level energies within an
ion stage were shifted by the same amount when implementing this procedure.
Of course, inaccuracies in the calculated ionization energies
are reflected in the individual level energies as well, but the line positions
are determined by taking the difference of energies within the same ion stage.
So some beneficial cancellation is expected in this regard when systematic
shifts are present within a given ion stage.
The NIST ionization-energy correction was applied to all of the models
considered in this work. An illustration of the ionization balance
that is obtained with these improved energies is provided
in Figure~\ref{fig:ionfrac_nd} for Nd at a mass density
of $\rho = 10^{-13}$~g/cm$^3$, corresponding to the ejecta density at
$\sim 1$~day after the merger (see Figure~3 in \citealt{rosswog14}).
This behavior is typical for all of the elements considered in this work.

\subsection{Variant Models}
\label{subsec:models_var}

In addition to calculating FR and SR models, two less-accurate (but faster
to compute)
FR models were generated in order to test the sensitivity of the macronovae 
emission to the quality of the atomic data. Configuration interaction (CI)
is a method to better describe the correlation between the bound electrons
of an atom or ion, and typically results in improved level energies
and oscillator strengths (for a more detailed explanation, see, for
example, \citealt{cowan,LANL_suite}). The use of CI is crucial for
obtaining reasonably accurate atomic structure data
for the near-neutral heavy elements considered here. However, due to
the smearing of lines caused by the large velocity gradients in the
ejecta, it is possible that differing amounts of CI could produce
similar spectra, which, if true, would provide more confidence in
the fidelity of the simulated spectra, at least from an atomic
physics perspective.

In order to test this concept, we generated two additional FR models
for Nd: one that includes CI between
only those basis states that arise from the same relativistic configuration
and one that includes CI between only those basis states that arise from the
same non-relativistic configuration. These models are referred to
here as ``FR-SCR" and ``FR-SCNR", respectively
(see \citealt{LANL_suite,fontes_cr16} for additional details). 
The FR-SCR model is less accurate than the FR-SCNR model, which is less accurate
than the FR model described above. All three FR models 
contain the same number of fine-structure levels, but their energies
differ due to the different CI treatments. Additionally,
each model contains a different number of lines, as displayed
in Table~\ref{tab:linecomp}. The variant models have fewer lines
than the baseline FR model, 
and those transitions that are common to the three models will
typically be described by different oscillator strengths.
\begin{table}[h]
\centering
\caption{\rm Number of lines per ion stage of neodymium for the
FR, FR-SCNR and FR-SCR models (see text).}
\vspace*{0.5\baselineskip}
\begin{tabular}{lrrr}
\hline
Ion stage   &   \multicolumn{3}{c}{\# of lines} \\
\cline{2-4}
            &    FR    &  FR-SCNR   & FR-SCR  \\
\hline
Nd {\sc i}  & 25,224,451 &  14,330,369  & 2,804,438  \\
Nd {\sc ii} &  3,958,977 &   3,222,445  &   783,275  \\
Nd {\sc iii}&    233,822 &     137,192  &    51,036  \\
Nd {\sc iv} &      5,784 &       5,393  &     2,051  \\
\hline
\end{tabular}
\label{tab:linecomp}
\end{table}

Unless otherwise noted, the tables and figures displayed
from this point forward refer to the baseline FR model.

\section{Opacities}
\label{sec:opac}

\begin{figure*}[h]
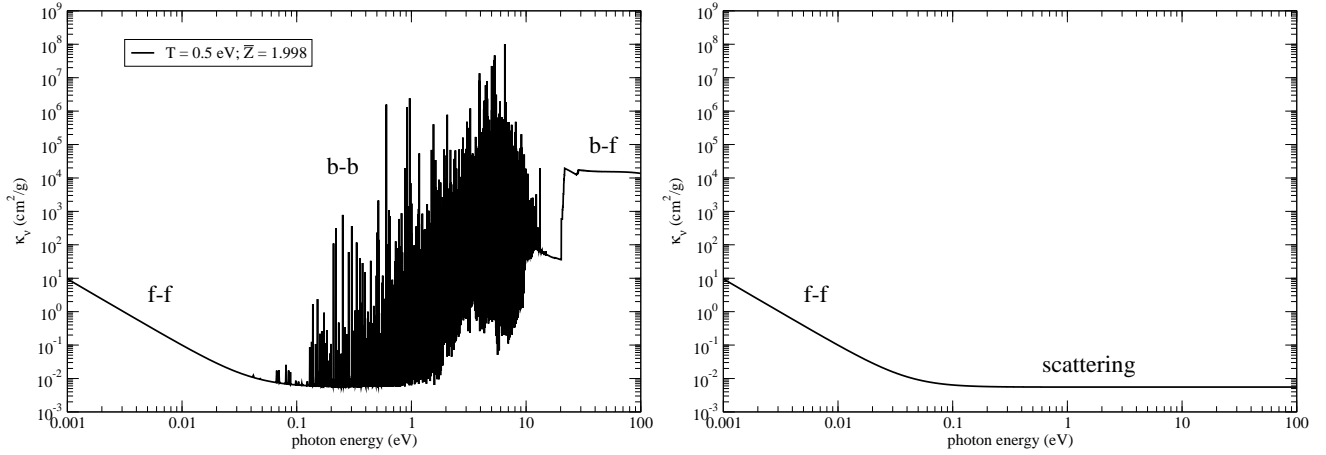

\includegraphics[clip=true,angle=0,width=1.0\columnwidth]
{opac_nd_kasen_fs_all_m13gcc_lte_nist_with_atwt_0.5ev_v2_PAPER.eps}
\includegraphics[clip=true,angle=0,width=1.0\columnwidth]
{opac_nd_kasen_fs_all_m13gcc_lte_nist_with_atwt_0.5ev_nobbbf_PAPER.eps}
\caption{
The LTE monochromatic opacity for neodymium at $T = 0.5$~eV and
$\rho = 10^{-13}$~g/cm$^3$. The left panel displays the complete opacity,
which includes the bound-bound, bound-free,
free-free and scattering contributions.
The right panel displays only the contributions due to free electrons,
i.e. the free-free and scattering contributions.
The average charge state for these conditions is listed in the legend
of the left panel.
}
\label{fig:opac_nd_1}
\end{figure*}

In order to illustrate the basic characteristics of the opacities used
in this study, the LTE monochromatic opacity for Nd is displayed in
Figure~\ref{fig:opac_nd_1}
for typical ejecta conditions of $T = 0.5$~eV and
$\rho = 10^{-13}$~g/cm$^3$.
The left panel displays the complete opacity, with all four
contributions (b-b, b-f, f-f and scattering), while the right panel
shows the contributions that arise only from free electrons (f-f and
scattering) in order to give some indication of the massive
differences that occur when the bound electrons are taken into
account.  The f-f and scattering contributions were obtained from the
simple, analytic formulas \citep{huebner} associated with Thomson and
Kramers, respectively. The gap between the b-b features and the onset
of the b-f edge occurring at $\sim 20$~eV is due to missing lines that
would be present if more excited configurations had been included in
the model. Our transport calculations in Section~\ref{sec:lc} assume
this gap is filled in and minimal transport occurs at these energies.
We note that a mean charge state of $\overline{Z} = 1.998$ is
obtained for these conditions, indicating that the opacity is
dominated by Nd~{\sc iii}.

In this example, the line-profile function in Equation~(\ref{opac_bb})
was chosen to be a standard Voigt profile, which is the convolution
of a Lorentz and Gaussian profiles, and takes the natural and
thermal Doppler widths as input for a given line. (We note
that the latter Doppler broadening is not the same as the previously discussed
line-smearing approach,
which is addressed below.) The inclusion of the line features dramatically
increases the opacity in the optical range
(1.65--3.26~eV or 0.751--0.380\,$\mu$m) by up to
eight orders of magnitude. The absorption in the near-infrared range below
1.65 eV (0.496--1.65~eV or 10.0--0.751\,$\mu$m) is also greatly increased,
by a few orders of
magnitude in this case, indicating that spectra would more likely be observed
in the mid-infrared range, at least for these specific conditions.

\subsection{Line-smeared opacities}
\label{sub:opac_lb}

\begin{figure*}[h]
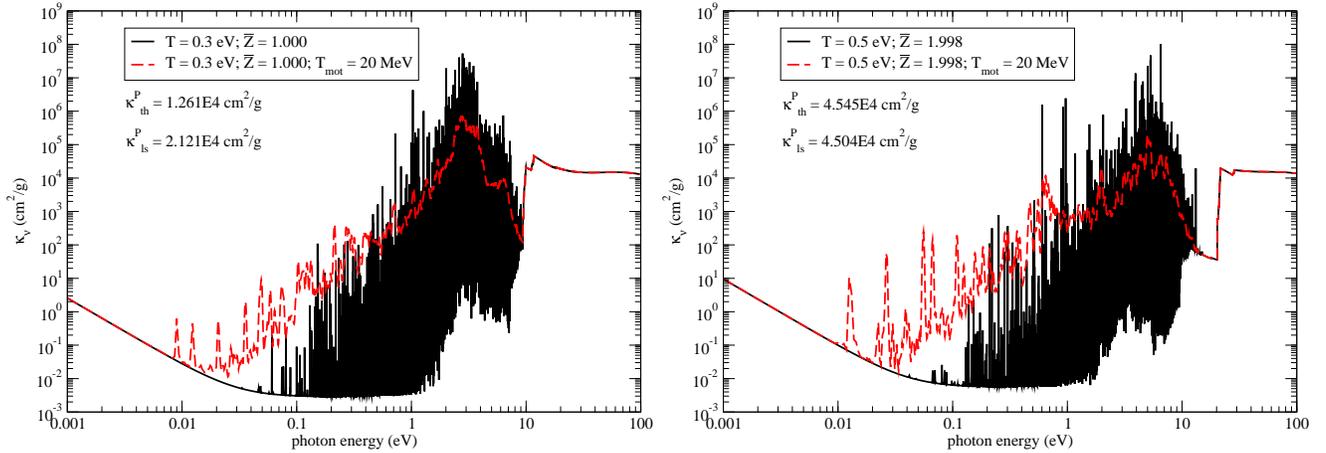

\includegraphics[clip=true,angle=0,width=1.0\columnwidth]
{opac_nd_kasen_fs_all_m13gcc_lte_nist_with_atwt_0.3ev_20MeVwid_v2_PAPER_LS.eps}
\includegraphics[clip=true,angle=0,width=1.0\columnwidth]
{opac_nd_kasen_fs_all_m13gcc_lte_nist_with_atwt_0.5ev_20MeVwid_v2_PAPER_LS.eps}
\caption{
The LTE monochromatic opacity for neodymium at $T = 0.3$~eV (left panel)
and $T = 0.5$~eV (right panel) for a mass density of
$\rho = 10^{-13}$~g/cm$^3$. The solid black curve in each panel includes
a bound-bound contribution that was calculated with a Voigt profile using
the appropriate natural width and thermal Doppler broadening for each line.
The dashed red curve is the same as the black curve, except that the
bound-bound contribution was calculated with smeared lines, using
an effective Doppler width evaluated at a characteristic temperature
of $T_{\rm ls} = 20$~MeV.
The average charge state for these conditions is listed in the legend
of each panel.
The Planck mean opacities obtained via integration of the thermally
broadened and line-smeared opacities are also provided.}
\label{fig:opac_nd_2}
\end{figure*}

As an example of the line-smeared opacities used in our light-curve
simulations, we present Figure~\ref{fig:opac_nd_2}.
The conditions from Figure~\ref{fig:opac_nd_1} ($T = 0.5$~eV and
$\rho = 10^{-13}$~g/cm$^3$) are repeated in the right panel, while a lower
temperature of $T = 0.3$~eV was chosen for the left panel, which is
dominated by contributions from Nd~{\sc ii} $(\overline{Z} = 1.000)$.
In this case, both the black solid curve and the red dashed curve
in each panel include b-b features that were calculated with
Voigt profiles that included the same set of natural widths, but different
Doppler widths. The solid curve represents the traditional thermal broadening
evaluated at the LTE temperature, while the dashed curve takes into account
the smearing of lines via a much higher effective
temperature ($T_{\rm ls} = 20$~MeV).
This temperature is obtained by assuming a velocity gradient
of $\Delta v/c \sim \Delta \nu/\nu \sim 0.01$, where $c$ is the
speed of light.  This gradient mimics the velocity gradient 
across a cell in our 2-dimensional light-curve models and is consistent 
with our energy group resolution in our transport calculations
(see Section~\ref{sec:lc}).
We substitute this gradient value into the standard expression for the Doppler
width of a transition, $i$, centered at frequency $\nu_i$, i.e.
\begin{equation}
\Delta \nu_i = \nu_i \left( \frac{2kT}{Mc^2}\right)^{1/2}
\label{doppler}
\end{equation}
to obtain an effective temperature, $T = T_{\rm ls}$, which is used
in place of the thermal temperature to smear the lines
in the ejecta. In the above expression,
$M = A/A_0$ is the mass of an atom of the element of interest,
$A$ is the atomic weight and $A_0$ is Avogadro's number.
Using a value of $A = 144$ for Nd, we arrive at an effective temperature
$T_{\rm ls} \sim 20$~MeV. This temperature was used to generate
line-smeared opacities for the three lanthanide elements considered
here (Ce, Nd, Sm), since they have similar atomic weights, while a
temperature of $T_{\rm ls} = 33$~MeV was chosen for U, based on the
linear scaling of the broadening temperature with atomic weight.
These effective temperatures were used only in the Doppler-width formula
appearing in Equation~(\ref{doppler}).

From Figure~\ref{fig:opac_nd_2}, we observe
that the effect of the line smearing is to smooth out the
jagged b-b features in the opacity, while preserving the area under the curve
because the integral of the Voigt profile over all frequencies is normalized
to one. Therefore, the line-smeared curves represent a kind of
average value of the b-b opacity at each photon energy. This behavior
is exhibited between energies of $\sim$~1--10 eV. On the other hand,
the area under the curves
at lower photon energies is strikingly different, with the
line-smeared curve actually exceeding the thermal curve by a significant
amount.
This difference indicates that the thermal curve contains very narrow lines
that are not fully resolved by the chosen grid spacing of photon energies
(from Equation~(\ref{doppler}), the thermal broadening decreases
as the photon energy decreases),
and so opacity is actually lost in that calculation due to this deficiency.
For the curves with line smearing, the lines are sufficiently
broad that they are adequately resolved and we obtain a faithful
representation of the frequency-dependent opacity within this approximation.
So, the use of line smearing in this manner provides
the additional (computational) advantage of being able to represent the
monochromatic opacities with a reasonable number of photon energy points for
the expected range of ejecta conditions, making possible the
generation of opacity tables that can be used in an efficient
look-up approach in our radiation transport simulations.

The amount of missing opacity that results from the lack of photon-energy
resolution can be roughly estimated by comparing
the Planck mean opacity, defined as
\begin{equation}
\kappa^{\rm P} \equiv \int_0^\infty B_\nu(T) \kappa_\nu' \,d\nu /
\int_0^\infty B_\nu(T) \,d\nu \,,
\label{planck}
\end{equation}
where $B_\nu(T)$ is the Planck function and $\kappa_\nu'$ indicates
that the scattering contribution is omitted from the
monochromatic opacity.
These mean values are also displayed in Figure~\ref{fig:opac_nd_2}, indicating
a 70\% difference at $T = 0.3$~eV, but almost no difference at
the higher temperature of $T = 0.5$~eV. The latter null result is partially due
to the fact that the peak of the Planck weighting function occurs at a photon
energy of $\sim 2.7\,T$ and the two monochromatic curves agree better
at higher photon energies. Despite this similarity in mean values,
a significant difference persists in the monochromatic opacities
at lower photon energies,
as can be seen in the right panel of Figure~\ref{fig:opac_nd_2}.

\begin{figure*}
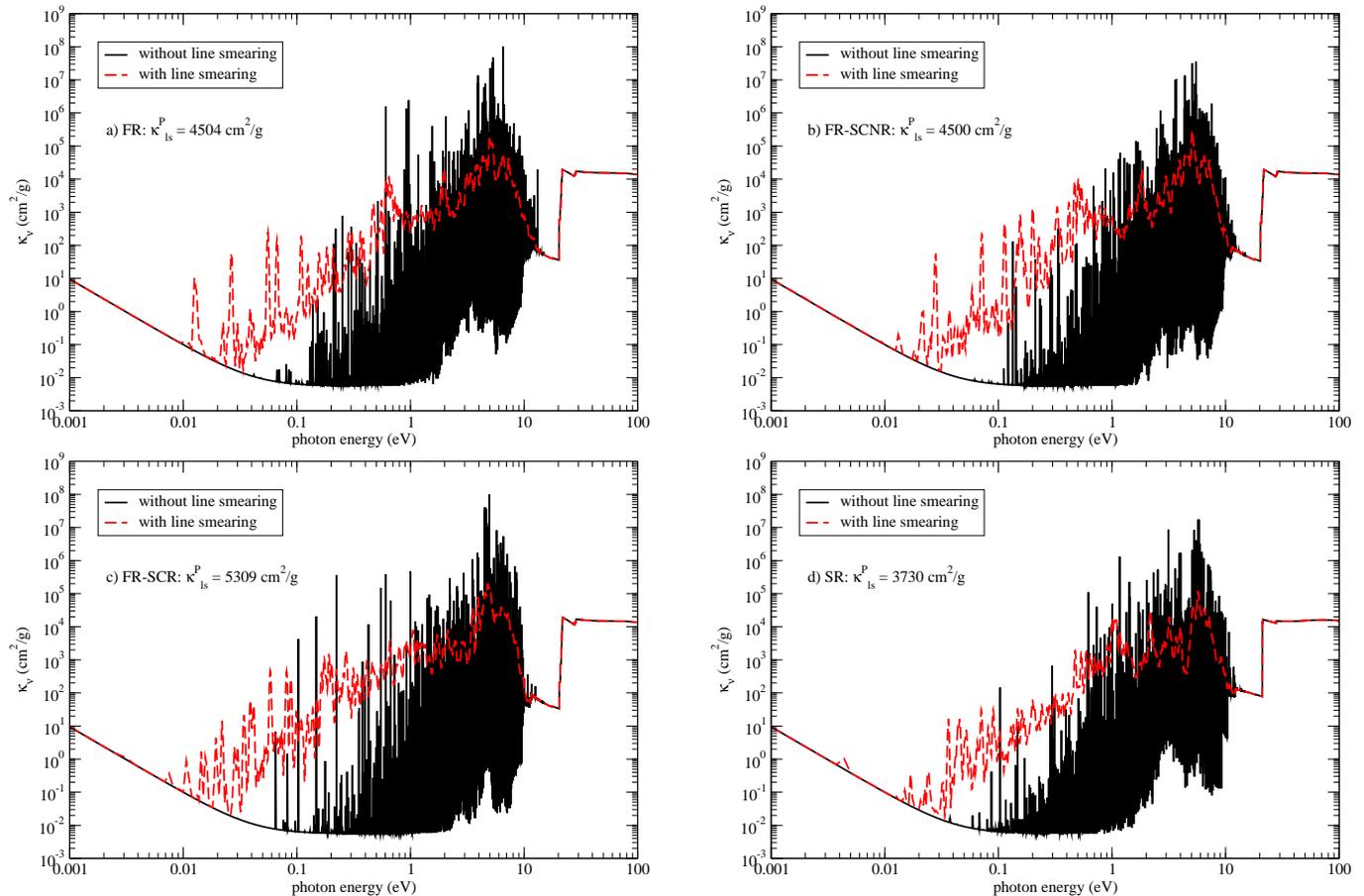

\includegraphics[clip=true,angle=0,width=1.0\columnwidth]
{opac_nd_kasen_fs_all_m13gcc_lte_nist_with_atwt_0.5ev_20MeVwid_v2_PAPER_2_LS.eps}
\hfill
\includegraphics[clip=true,angle=0,width=1.0\columnwidth]
{opac_nd_kasen_fs_all_m13gcc_lte_nist_with_atwt_0.5ev_20MeVwid_SCNR_v2_PAPER_2_LS.eps}
\includegraphics[clip=true,angle=0,width=1.0\columnwidth]
{opac_nd_kasen_fs_all_m13gcc_lte_nist_with_atwt_0.5ev_20MeVwid_SCR_v2_PAPER_2_LS.eps}
\hfill
\includegraphics[clip=true,angle=0,width=1.0\columnwidth]
{opac_nd_kasen_fs_all_m13gcc_lte_nist_with_atwt_0.5ev_20MeVwid_SR_v2_PAPER_2_LS.eps}
\caption{
The LTE monochromatic opacity for neodymium at $T = 0.5$~eV
and $\rho = 10^{-13}$~g/cm$^3$ using four different models
described in the text:
a) FR, b) FR-SCNR, c) FR-SCR, and d) SR.
The solid black curve in each panel includes
a bound-bound contribution that was calculated with a Voigt profile using
the appropriate natural width and thermal Doppler broadening for each line.
The dashed red curve is the same as the black curve, except that the
bound-bound contribution was calculated with smeared lines, using
an effective Doppler width evaluated at a characteristic temperature
of $T_{\rm ls} = 20$~MeV.
The Planck mean opacity obtained via integration of the line-smeared
opacity is also listed in each panel.
}
\label{fig:opac_nd_3}
\end{figure*}

A second Nd opacity example is provided in Figure~\ref{fig:opac_nd_3},
with the conditions ($T = 0.5$~eV, $\rho = 10^{-13}$~g/cm$^3$) being the
same as those given in the right panel Figure~\ref{fig:opac_nd_2}.
Monochromatic opacities are presented for the four models described
in Section~\ref{sec:atomic}: FR, FR-SCNR, FR-SCR, SR. The models produce
qualitatively similar results, but there are visible quantitative differences,
exemplified by the Planck mean opacities displayed in each panel.
The SR model produces the smallest mean value, given by 3730~cm$^2$/g,
while the least accurate FR-SCR model has a value of 5309~cm$^2$/g,
resulting in a difference of 42\%. The most accurate FR model produces
an intermediate value of 4504~cm$^2$/g, with the FR-SCNR yielding
a similar result. These differences allow us to test the sensitivity
of the light curves to changes in the underlying atomic physics that
is used to construct the opacity.

\begin{figure*}
\includegraphics[clip=true,angle=270,width=1.0\columnwidth]
{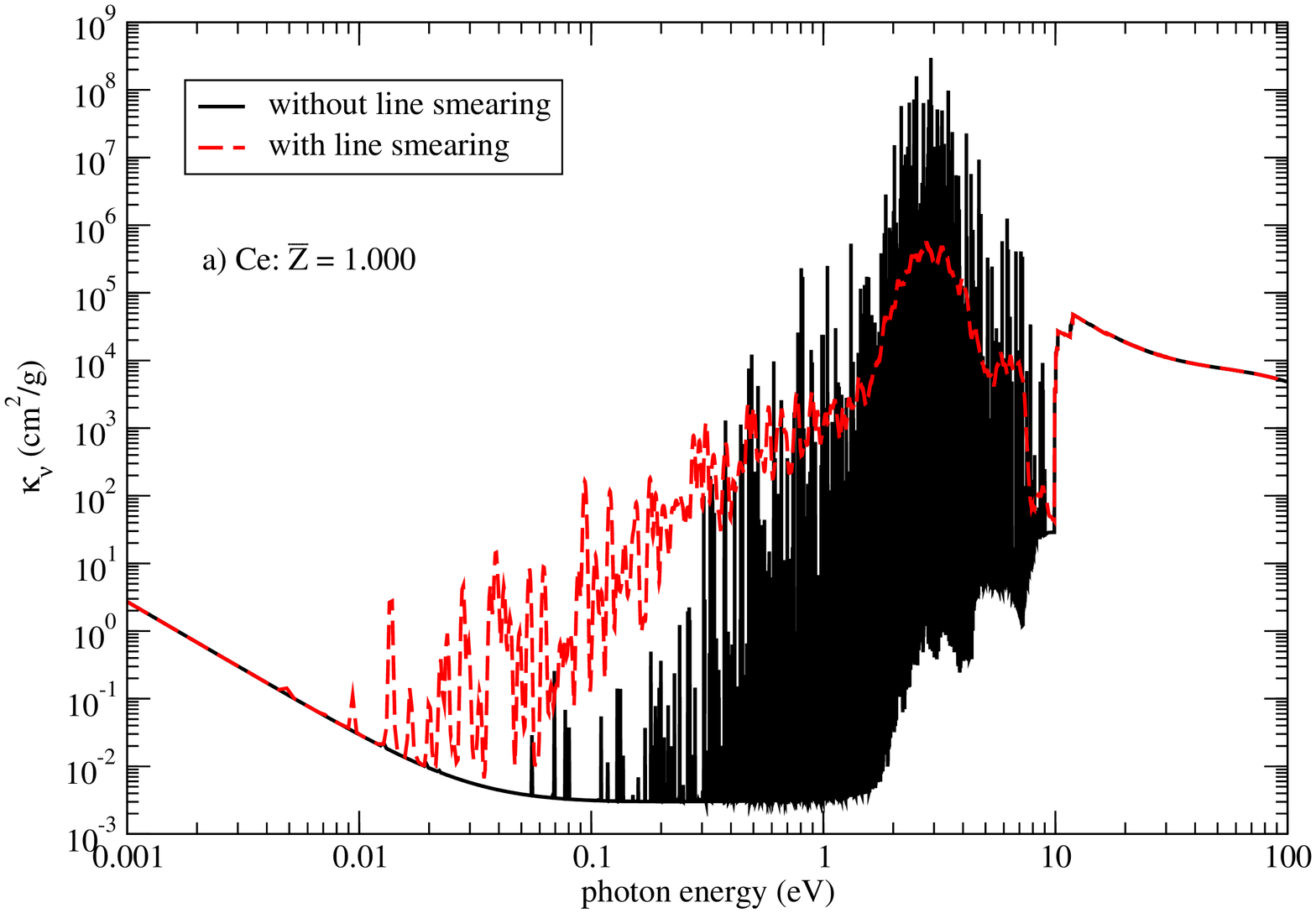}
\hfill
\includegraphics[clip=true,angle=270,width=1.0\columnwidth]
{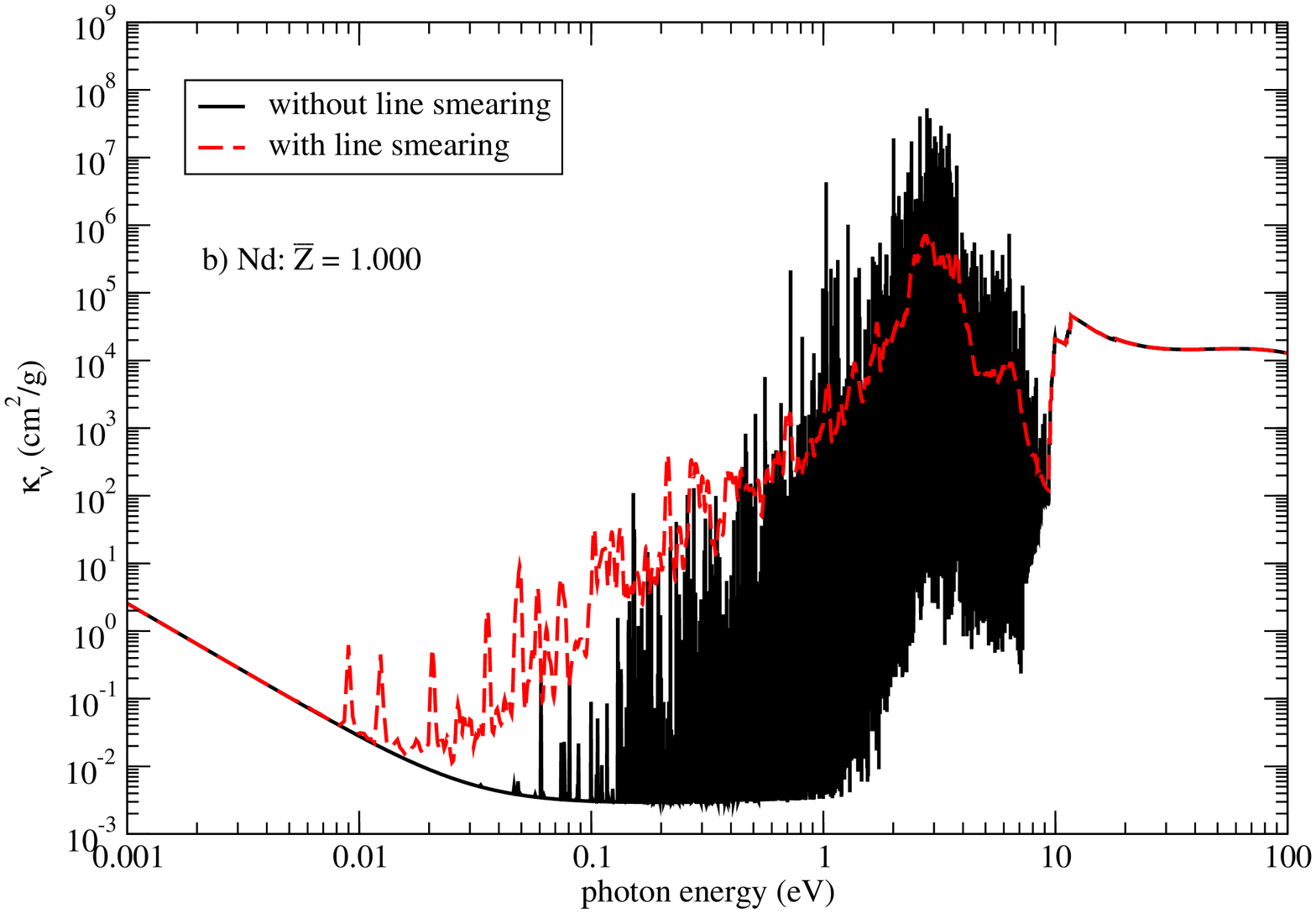}
\includegraphics[clip=true,angle=270,width=1.0\columnwidth]
{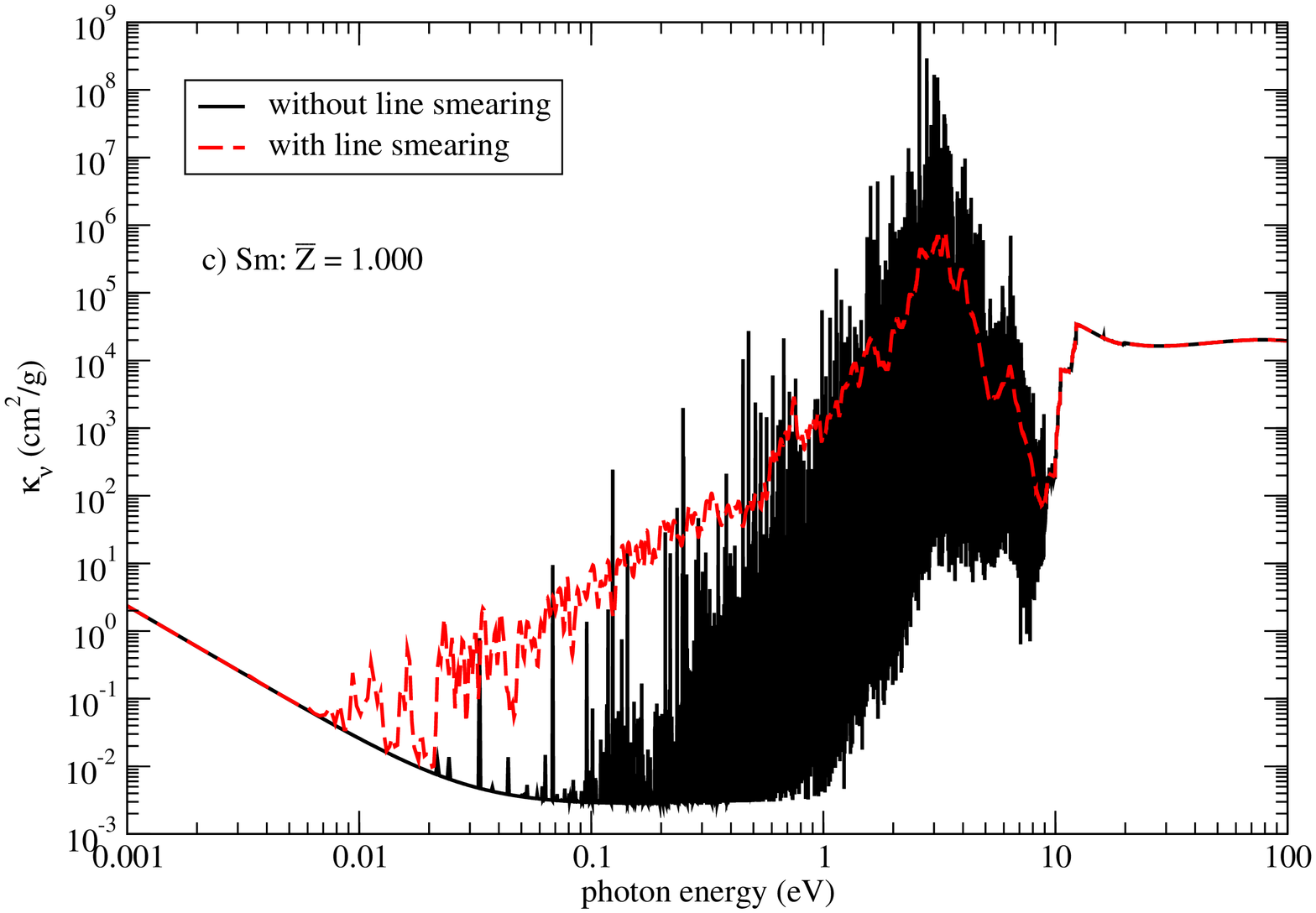}
\hfill
\includegraphics[clip=true,angle=270,width=1.0\columnwidth]
{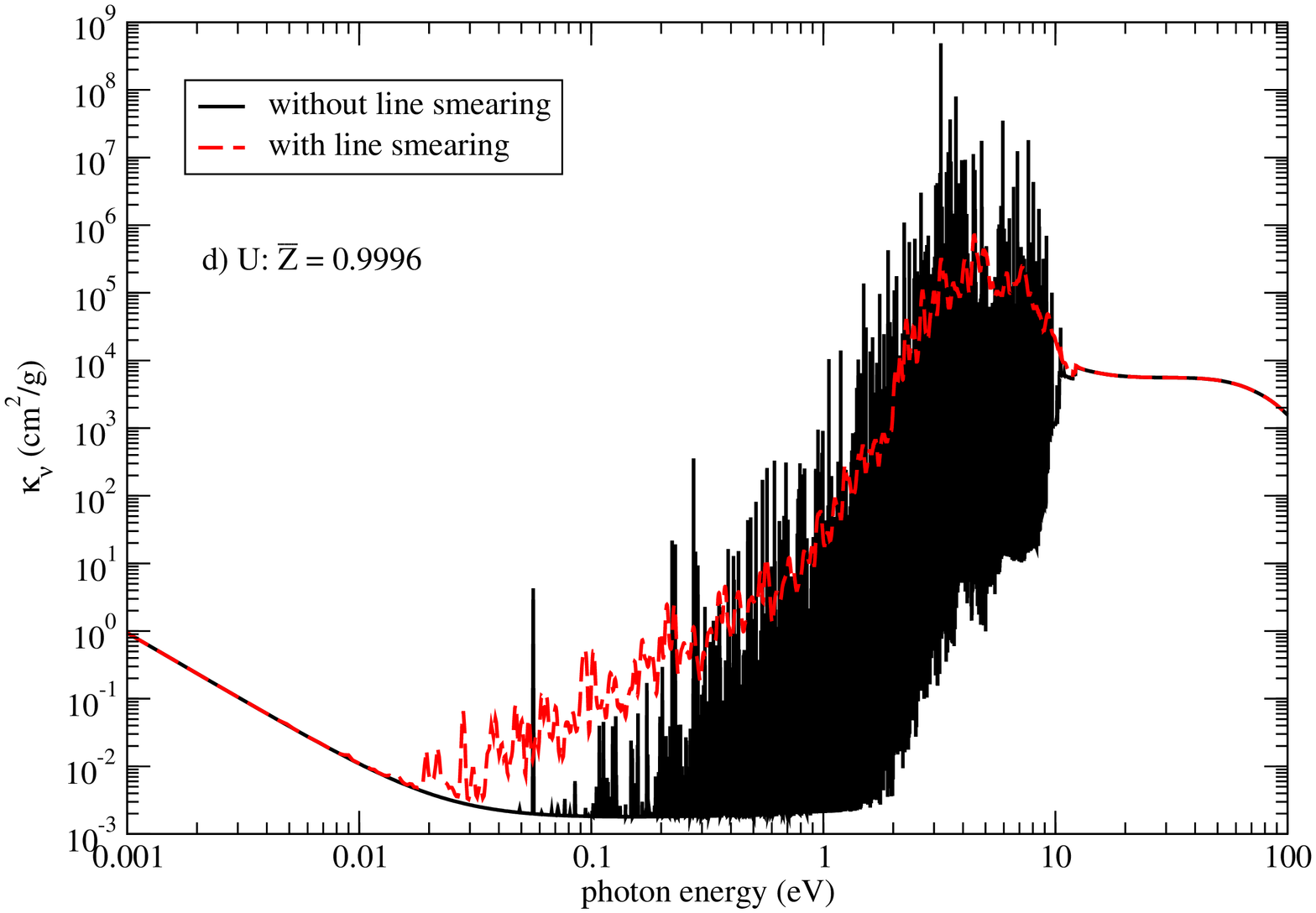}
\caption{
The LTE monochromatic opacity for the four elements considered in
this work at $T = 0.3$~eV
and $\rho = 10^{-13}$~g/cm$^3$:
a) cerium, b) neodymium, c) samarium, and d) uranium.
The solid black curve in each panel includes
a bound-bound contribution that was calculated with a Voigt profile using
the appropriate natural width and thermal Doppler broadening for each line.
The dashed red curve is the same as the black curve, except that the
bound-bound contribution was calculated with smeared lines, using
an effective Doppler width evaluated at a characteristic temperature
of $T_{\rm ls} = 20$~MeV for the three lanthanides and $T_{\rm ls} = 33$~MeV
for uranium.
The mean charge is also displayed in each panel.
}
\label{fig:opac_all}
\end{figure*}

As a final illustration, we present in Figure~\ref{fig:opac_all}
the monochromatic opacities for all four elements in this study
at conditions of $T = 0.3$~eV, $\rho = 10^{-13}$~g/cm$^3$.
The opacity for each element displays the same basic trends. Each element
has $\overline{Z} \approx 1$ at these conditions due to the similarity in
ionization energies, as displayed in Table~\ref{tab:ionpot}.
Therefore, the bound-bound contribution to the opacity is dominated by lines associated
with the singly ionized stage. The absorption peaks
at a photon energy of $\sim 3$~eV, with decreasing opacity as the photon energy decreases
(or as wavelength increases). However, the detailed line structure is different
for each element, with the complexity increasing as the ground configuration
of the dominant ion stage moves closer to a half-filled $f$-shell, for
which the complexity of angular momentum coupling is maximal.
In all cases, a significant amount
of opacity is missing at lower wavelengths when thermally broadened lines 
(black, solid curves) are used, but is captured when smeared lines
are considered (red, dashed curves).

\subsection{Justification for line-smeared opacities}
\label{sub:just}

The issue of which method to choose for calculating opacities in
a moving medium is not straightforward.
For circumstances with a lack of strong line overlap and insignificant
fluorescence, the Sobolev approach is well founded. However, the justification
for reducing the optical depth under the Sobolev approximation is less obvious
if a photon absorbed by a line resonance often leaves with a significantly
different energy, i.e. fluorescence.
This issue is similar to the discussion of whether to use Rosseland or
Planck opacities for a given problem, e.g.~\citep{pomraning71}.
Typically, it is assumed that if one is focused on the line opacities
(e.g. line emission), the Planck opacity is better. However, if one is focused
on radiative heat transfer, the opacity between the lines where radiation can
flow more freely is most important. As such, most calculations that focus on
heat flow use the Rosseland opacity, e.g.~\citep{ganapol87}. Similarly, the
expansion-opacity formulation
is an approximation for estimating the opacity through a medium with a velocity
gradient~\citep{karp77}. This method uses the Sobelev line-strength
approximation~\citep{sobolev60} for radiation flow through velocity gradients.
Type~Ia supernovae modeled with this expansion opacity can reproduce bolometric
light curves, but the resulting spectra are too blue~\citep{pintoeast00}.
To address this discrepancy, \citet{pintoeast00} argued that a correction
factor was needed to
include fluorescence in the lines, neglected in the Sobolev
approximation~\citep{sobolev60}. The correction factor is chosen to match the
Type~Ia light curves.

\begin{figure*}
\includegraphics[clip=true,angle=0,width=1.0\columnwidth]
{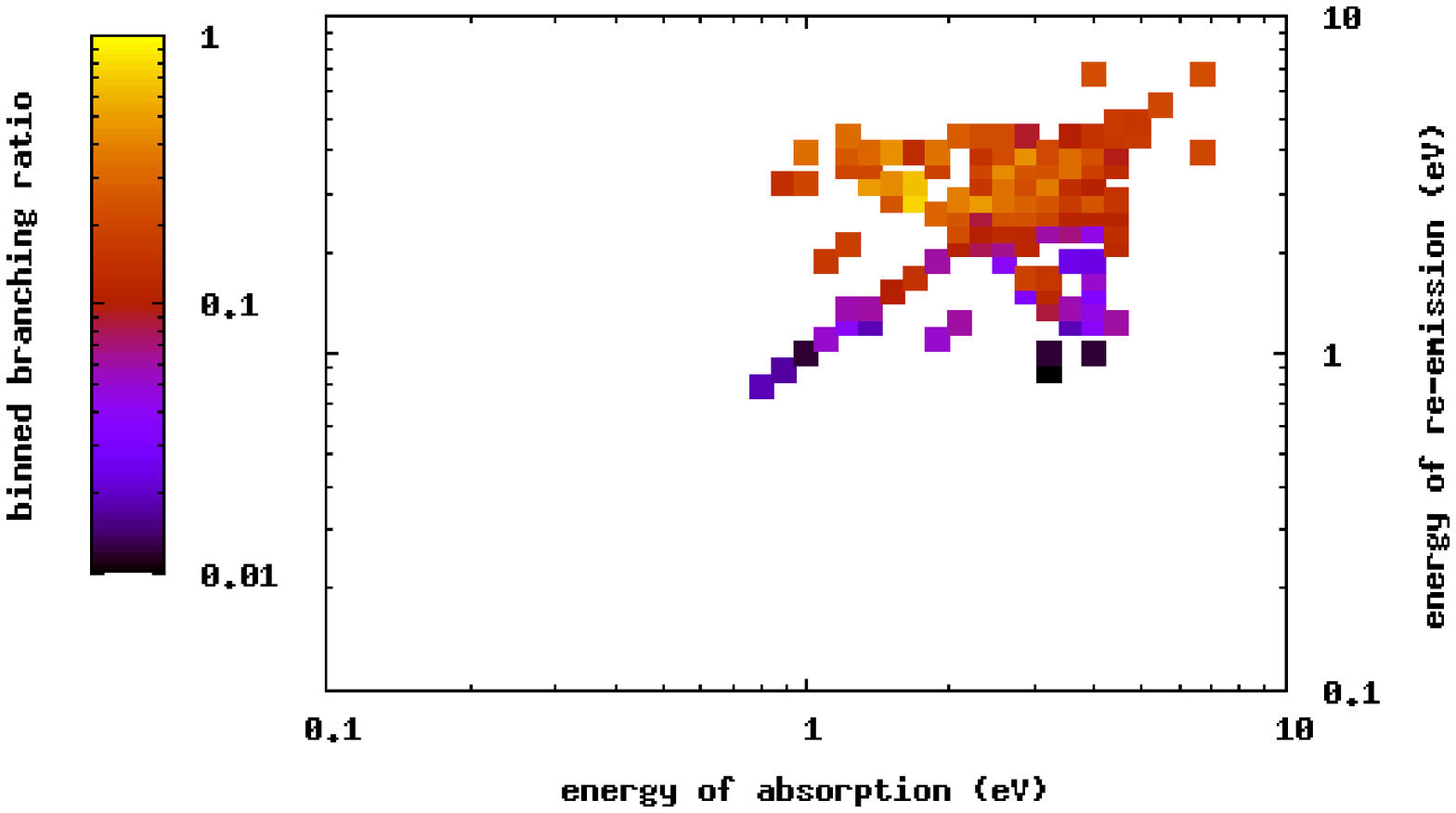}
\includegraphics[clip=true,angle=0,width=1.0\columnwidth]
{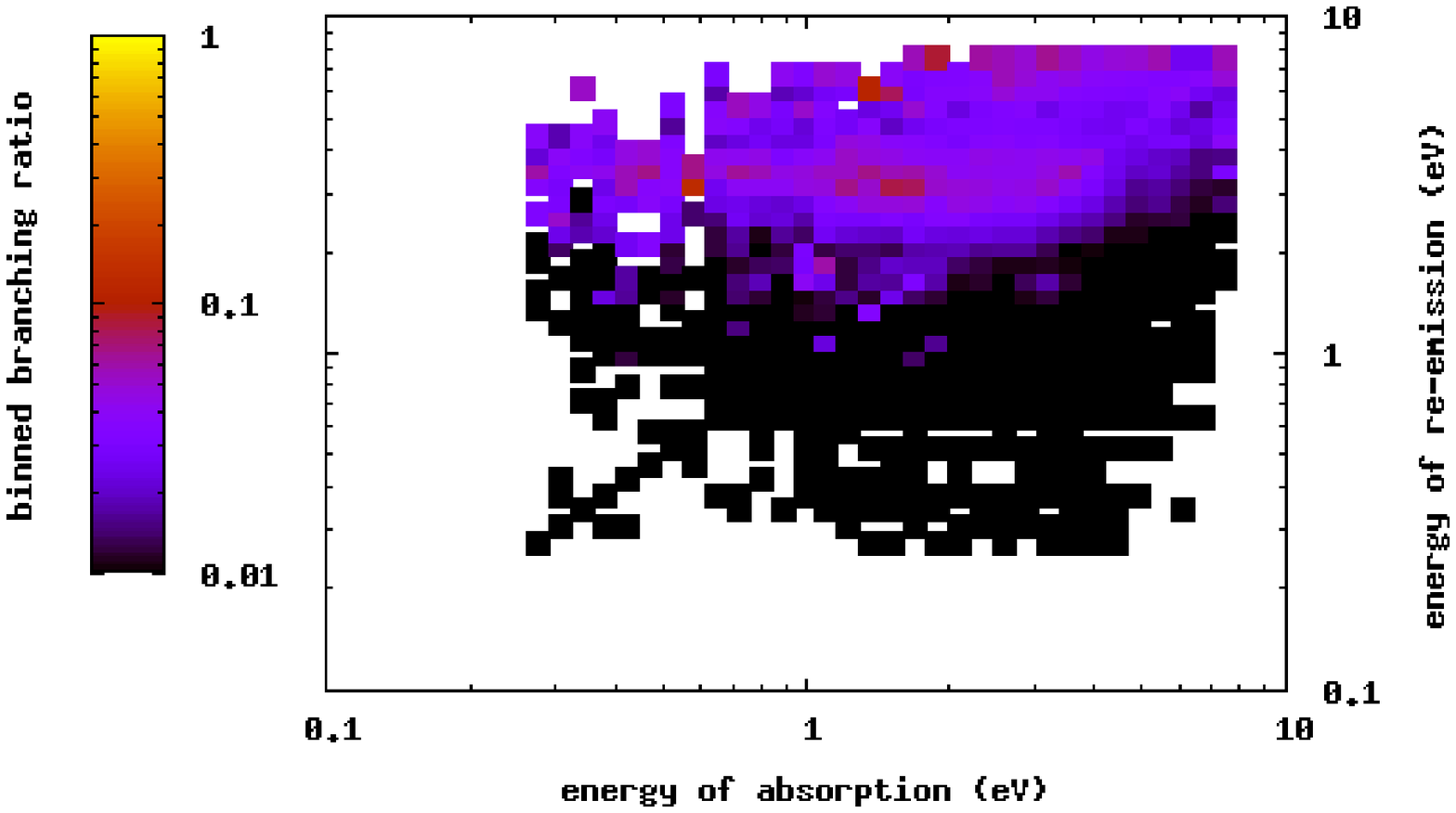}
\caption{
Radiative-decay branching ratios for transitions of various strengths
in Nd~{\sc ii}.
The left panel displays branching ratios for the strongest set
of absorption lines with oscillator strengths that satisfy $f_{lu} > 0.1$.
The right panel displays branching ratios for a weaker set of absorption lines
with oscillator strengths in the range $0.01 < f_{lu} \le 0.1$.
}
\label{fig:brxrad}
\end{figure*}

The case has been made for relatively high fluorescence in Type~Ia supernovae
by \citet{pintoeast00} and by \citet{kasen06_2}.
A similar effect has been suggested by \citet{kasen13} for NS ejecta.
As previously mentioned, the lanthanides and actinides are expected to
provide the dominant contribution to the ejecta opacity. The complex
energy-level structure of these elements supports the possibility
of high fluorescence. As an illustrative example, we consider
Figure~\ref{fig:brxrad}, which displays the radiative-decay branching ratios for
transitions of various strengths in Nd~{\sc ii}.
The branching ratio for radiative decay from an upper level $u$ to a lower
level $l$ is given by the standard expression
\begin{equation}
B_{ul} = \frac{A_{ul}}{\sum_k A_{uk}} \,,
\end{equation}
where $A_{ul}$ is the Einstein coefficient for spontaneous radiative decay
from level $u$ to $l$. The summation in the denominator
includes all allowed decay channels from the upper level $u$.
The two panels in Figure~\ref{fig:brxrad} display the branching
ratio in color-contour format, with the intensity representing the
probability that a photon absorbed with a particular energy given on the $x$
axis will be re-emitted at an energy given on the $y$ axis
(assuming no other processes, such as electron collisions, play a
role in destroying the upper level $u$).
The left panel displays the branching ratio for the strongest set of
absorption lines in our model for Nd~{\sc ii}, i.e. $f_{lu} > 0.1$,
where $f_{lu}$ is the absorption oscillator strength for a photon
absorption from level $l$ to level $u$. (Here, were employ the
dual-index form of the oscillator strength, rather than the
single-index form displayed in Equation~(\ref{opac_bb}).)
The right panel displays the branching ratio for a weaker
set of Nd~{\sc ii} lines in the adjacent range given by
$0.01 < f_{lu} \le 0.1$.
The color contours were obtained by averaging the pointwise $B_{ul}$ values
over logarithmically spaced bins in the $x$ and $y$ directions.
Branching ratios that appear on the $x=y$ diagonal line correspond to
scattering, i.e. re-emission at the same energy possessed by the absorbed
photon, while the off-diagonal values represent fluorescence.

The strongest set of lines considered in the left panel displays a mixture
of both fluoresence and scattering.
These lines typically appear as strong absorption features
at optical and higher photon energies, where the monochromatic opacity peaks
for representative macronova conditions, such as those considered in
Figures~\ref{fig:opac_nd_1}--\ref{fig:opac_all}.
The weaker set of lines considered in the right panel displays a significant
amount of fluorescence, with the re-emitted energy spread across a much
broader range.
There are significantly more options for absorption and re-emission
in this set due to photo-absorption from the numerous excited levels
that lie above the ground (and lowest-lying excited) levels, as well
as due to the quantum selection rules that determine which transitions
are allowed. From this type of study, we find that strong fluorescence,
spread over a broad range of re-emission channels, persists for progressively
weaker absorption lines. A subset of these weaker, fluorescing lines permit
radiation to escape the ejecta, which produces the light curve.

Of course, the situation is more complicated than indicated by the above
illustrative example. A more thorough study should take
into account the actual population of the lower level for each transition
(see the factor of $N_i$ in Equation~(\ref{opac_bb})) at a given
temperature and density. Furthermore, collisional destruction
of the upper level, $u$, may be pertinent. Such issues have been
studied, for example, in the context of the expansion-opacity
formalism for Type~Ia supernovae
by \citet{pintoeast00} in their Figure~4, which displays the cascade
matrix for a particular set of conditions.
Nevertheless, our branching-ratio example suggests that
fluorescence is of primary importance in the lanthanide (and actinide)
contribution to the ejecta opacity.

We explore the consequences of the assumption of strong fluorescence
for macronovae in the present work using our line-smeared approach.
Our method represents an alternative
transport approximation that puts a primary emphasis on the fluorescence
phenomena compared to an expansion opacity with a calibrated correction factor.
For Type~Ia spectra, the higher opacity in the line-smeared type of
calculation moves the photosphere outward and the lower effective temperature
produces a better match with observation compared to
the expanding-photosphere method if the fluorescence correction is
omitted in the latter case.
Similarly, our approach produces significant differences
in the macronova light curves compared to those produced via
the opacity-expansion formalism.
The expanding-photosphere versus the full-line opacity methods represent two
possible options for calculating the supernova opacity, but, just like the
Rosseland and Planck mean opacities, these methods are approximations.
Because of the importance of fluorescence, we believe the method employed
here is more appropriate for supernova and macronova light-curve
calculations.

In support of our approach, we demonstrate the effectiveness of our
line-smeared opacities in modeling a simplified W7 problem for a pure-Fe
outflow, but with the same radioactive heating as in the standard W7 case.
This example uses the {\tt SuperNu} radiation
transport code, which has been previously shown to produce light curves
that are in good agreement with those generated with the PHOENIX code
\citep{wollaeger14} for the W7 Type~Ia light-curve problem, as well as an
observation of an abnormal Ia supernova
\citep{vanrossum16}. (The discrepancy in peak
luminosity described in \citet{wollaeger14} was reported to be 10-15\%, but has
subsequently been reduced to ~5--10\%, due to differences in gamma-ray
energetics (van Rossum \& Wollaeger, unpublished).)
Those {\tt SuperNu} W7 calculations
employed a Kurucz line list that was ``regrouped" for computational
efficiency. This regrouping is similar to our line-smeared approach,
i.e. individual lines are combined in such a way that the resulting opacity has
the same area under the curve as the ungrouped opacity. As mentioned
previously, an advantage of this type of
approach over the Sobolev line-expansion method is that
it does not require a fluorescence correction to obtain physically
reasonable light curves.

\begin{figure*}
\includegraphics[clip=true,angle=0,width=1.0\columnwidth]
{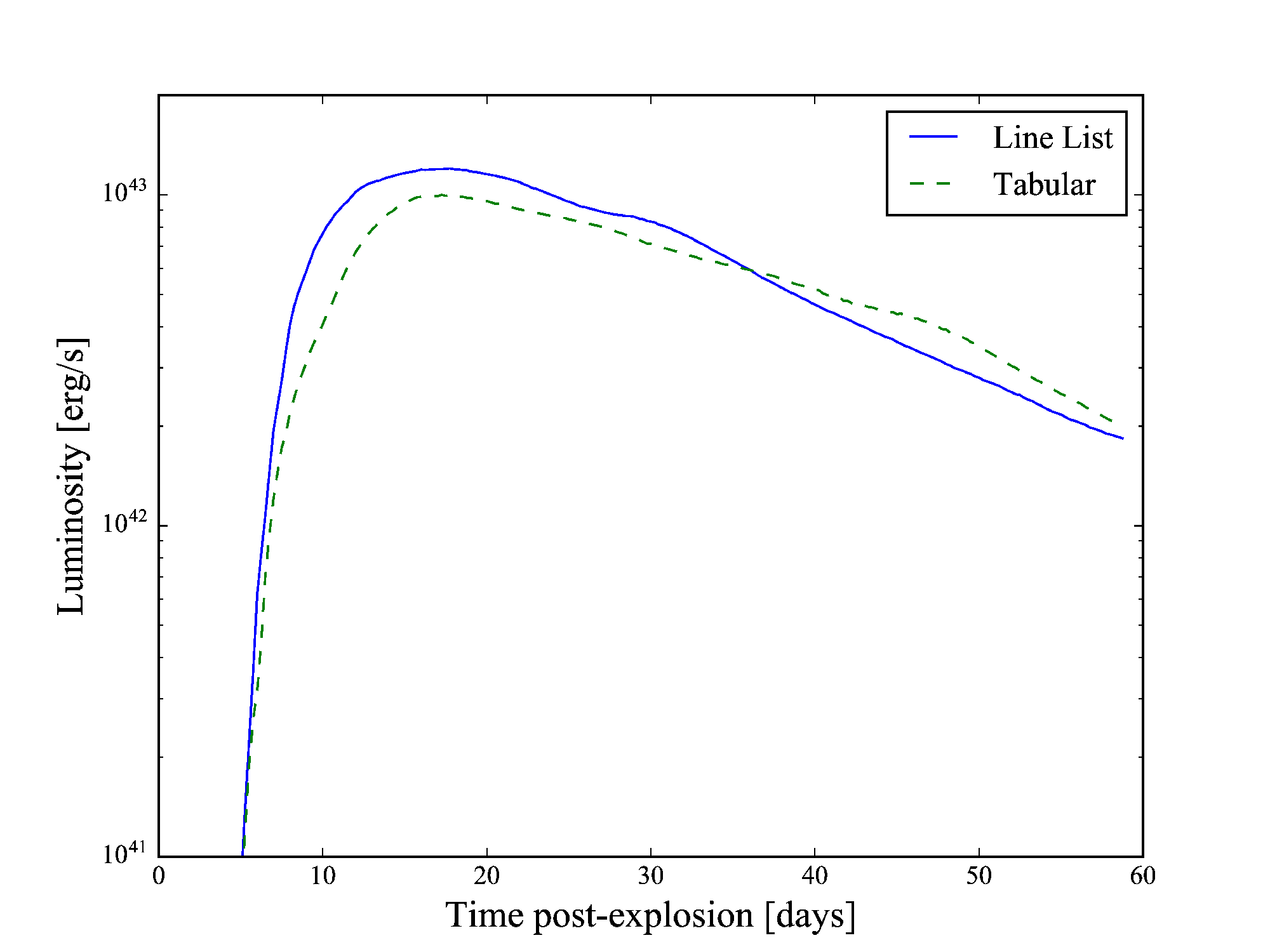}
\includegraphics[clip=true,angle=0,width=1.0\columnwidth]
{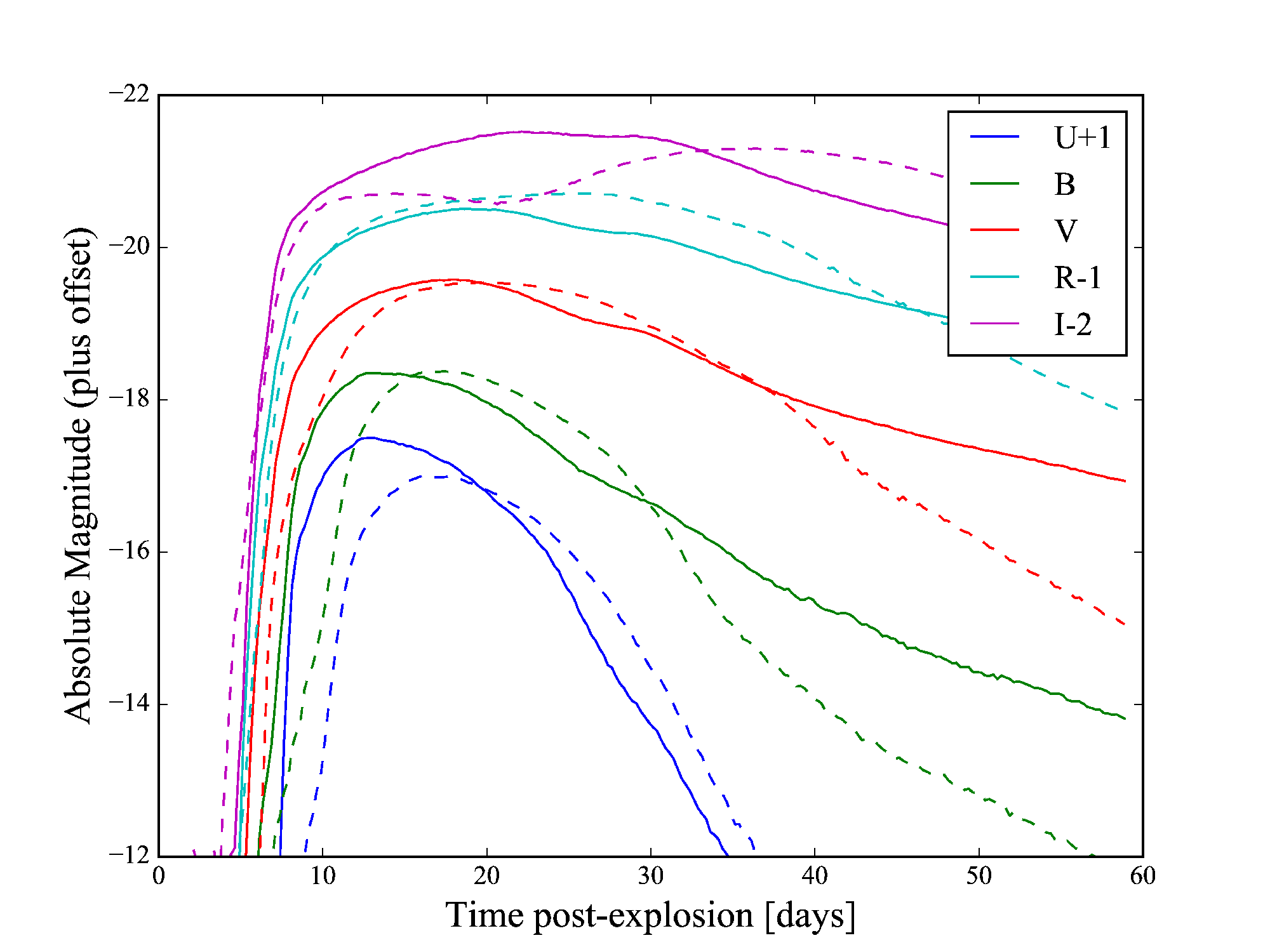}
\caption{
Left panel: the total W7 luminosity calculated with
{\tt SuperNu} using opacities obtained
from the Kurucz line list (solid, blue curve) and
from our tabular, line-smeared approach (dashed, green curve).
Right panel: absolute magnitude versus time since explosion for
the U, B, V, R and I bands. The solid curves represent the line-list
opacities and the dashed curves represent the present line-smeared
opacities. For clarity, the U-, R- and I-band curves have
been shifted by +1, -1 and -2, respectively.}
\label{fig:w7test}
\end{figure*}

In Figure~\ref{fig:w7test}, we compare
the total luminosity and UBVRI AB magnitudes for the W7 problem
obtained with opacities from the Kurucz line list and from our
line-smeared approach. The agreement is quite good for the total
luminosity, within $\sim 20$\% near and after peak luminosity.
The broadband light curves indicate that the bulk emission is in a similar
location in frequency around peak bolometric luminosity. The biggest discrepancy
occurs in the I band due to the dip in the tabular result. At the time of peak
total luminosity, the differences in the broadband luminosities (converting
from magnitude to equivalent luminosity) are:
$\sim 18$\% in the U band,
$\sim 15$\% in the B band,
$\sim 7$\% in the V band,
$\sim 15$\% in the R band and
$\sim 90$\% in the I band (due to the dip).

Next, we present a sample comparison of single-element
opacities generated with the
expansion-opacity and line-smeared methods for typical macronova
conditions. As mentioned previously,
the expansion-opacity method \citep{sobolev60,castor74,karp77} employed
by \citet{kasen13,barnes13} to simulate NSM light curves
applies to the bound-bound contribution to the
opacity and involves a discrete sum over all lines. The approach relies on the
assumption of a homologous expansion and is characterized by an expansion
time, $t_{\rm exp}$. The relevant wavelength range is divided into bins denoted
by index $j$, $\Delta\lambda_j$, and all lines within a bin are summed
to obtain the opacity for that range. The expression for the opacity
associated with bin $j$ is given by
\begin{equation}
\kappa^{\rm b-b}_{\rm exp}(\lambda_j) = \frac{1}{\rho c t_{\rm exp}}
\sum_i \frac{\lambda_i}{\Delta\lambda_j} (1 - e^{-\tau_i}) \,,
\label{opac_bb_exp}
\end{equation}
where $t_{\rm exp}$ is the time since mass ejection, the summation index $i$
extends over
all possible bound-bound transitions, $\lambda_i$ is the associated rest
wavelength, and $\tau_i$ is the corresponding Sobolev optical depth, i.e.
\begin{equation} 
\tau_i = \frac{\pi e^2}{m_e c}\, N_i\, |f_i|\, t_{\rm exp}\, \lambda_i \,.
\end{equation}

\begin{figure}
\includegraphics[clip=true,angle=0,width=1.0\columnwidth]
{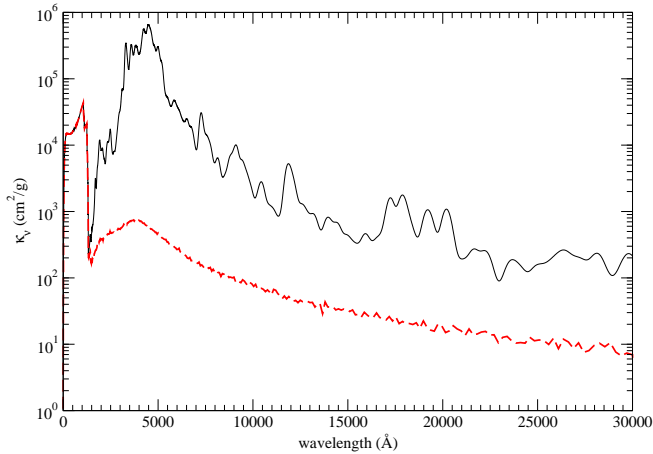}
\caption{
The LTE monochromatic opacity for neodymium at $T = 4,000$~K ($0.345$~eV)
and $\rho = 10^{-13}$~g/cm$^3$ using the line-smeared
(black solid curve) and Sobolev expansion (red dashed curve) methods.
The expansion opacity was calculated at a time since ejection of
$t_{\rm exp} = 1$~day and wavelength binning of $\Delta \lambda = 0.01 \lambda$.
The line-smeared method produces 
opacities that are typically one to two orders of magnitude higher than 
the expansion opacity.}
\label{fig:opac_nd_4}
\end{figure}

In Figure~\ref{fig:opac_nd_4}, we present the monochromatic opacity of Nd
at $T = $ 4,000~K ($0.345$~eV) and $\rho = 10^{-13}$~g/cm$^3$, generated
with the FR model. There are two curves, one representing
the bound-bound contribution obtained from the current
line-smeared approach, Equation~(\ref{opac_bb}),
and the other from the expansion-opacity method, Equation~(\ref{opac_bb_exp}).
(For completeness, we mention that the two corresponding curves generated with
the SR model (not shown) are quantitatively similar to the two FR curves
displayed in this figure.)
The opacity is plotted versus wavelength instead of energy,
and the conditions were chosen in order to facilitate direct comparison
with Figure~8 of \citet{kasen13}. The line-smeared approach produces
an opacity that is one or more orders of magnitude greater than the expansion
method over much of the wavelength range. As a consequence, the
luminosity will be significantly diminished when line-smeared
opacities are used to simulate light curves, compared to expansion values.

We note that the expansion opacity
in Figure~\ref{fig:opac_nd_4} is qualitatively similar to that
displayed in Figure~8 of \citet{kasen13}, with the peak value of the
b-b contribution occurring at $\sim$5,000~\AA\ and monotonically
decreasing at higher wavelengths.
However, the peak value is about three times larger in the present case,
providing a rough measure of the uncertainty in current opacity
calculations as they pertain to macronova conditions.
This discrepancy is somewhat surprising due to the fact that the same list
of configurations, resulting in the same number of lines (see the
Nd data listed in Table~\ref{tab:configs}), was used in both cases.
The differences are perhaps an indication of how difficult it
is to perform accurate
atomic structure calculations for such complicated atoms and ions.
Qualitative differences occur at the higher wavelengths, where the
present curve is significantly smoother and passes through
the relatively large peak-to-valley variations
displayed by the curve of \citet{kasen13}. We were able to obtain
such large oscillations in our calculations by employing a linearly
spaced wavelength grid, rather than the logarithmically spaced grid
obtained from the prescription $\Delta \lambda = 0.01 \lambda$.

\subsection{Opacity tables}
\label{sub:opac_tab}

In order to perform radiation-transport calculations in an efficient
manner, opacity tables were generated for the four elements of interest
using prescribed temperature and density grids that span
the range of conditions of interest.
The temperature grid consists of 27 values (in eV):
0.01, 0.07, 0.1, 0.14, 0.17, 0.2, 0.22, 0.24, 0.27, 0.3, 0.34, 0.4, 0.5, 0.6,
0.7, 0.8, 0.9, 1.0, 1.2, 1.5, 2.0, 2.5, 3.0, 3.5, 4.0, 4.5, and 5.0.
Specific temperature values are also indicated by circles in
the ionization balance plot of Figure~\ref{fig:ionfrac_nd}.
The density grid contains 17 values ranging from 10$^{-20}$
to 10$^{-4}$~g/cm$^3$, with one value per decade.

\section{Simulated Light Curves}
\label{sec:lc}

To provide a first look at how these different opacities affect the
macronova emission, we simulate a series of spectra and light curves
for our different opacities.  We use a simplified ray-trace transport
code to calculate the macronova light curves for different opacities
and atomic physics models, different merger models, and different
viewing angles.

We consider three merger models from \cite{rosswog14} and
\cite{grossman14} with the following binary masses (see
Table~\ref{tab:nsmerger}): $1.4+1.4\,{\rm M_\odot}$, $1.8+1.2\,{\rm
  M_\odot}$ and $1.6+1.2\,{\rm M_\odot}$ $18-25$\,ms after the onset
of the merger process.  For these three models, we consider only tidal
ejecta material with velocities above the escape velocity.  The ejecta
masses are 0.022, 0.043 and 0.041\,M$_\odot$, and the energies are
$1\times10^{50}$\,erg, $3.6\times10^{50}$\,erg and
$2\times10^{50}$\,erg for, respectively, the $1.4+1.4\,{\rm M_\odot}$,
$1.6+1.2\,{\rm M_\odot}$ and $1.8+1.2\,{\rm M_\odot}$ models.  We 
stress that these are just one set of merger models, and the assessment
of exact ejecta 
properties from different mergers is still an area of active research.  In each
model, the particles in the smooth particle hydrodynamics calculations
carried information of the particle positions at the end of the
calculation as well as the masses and velocities of the ejecta.
Assuming the ejecta are ballistic at the end of the calculations
(\cite{rosswog14} demonstrated the homologous expansion of this
ejecta), we can use this particle information to calculate the
3-dimensional distribution of matter.

\begin{table}[h]
\centering
\caption{Properties of our merger models.}
\vspace*{0.5\baselineskip}
\begin{tabular}{lcc}
\hline
Masses (M$_\odot$) & M$_{\rm ejecta}$ (M$_\odot$) & Energy$_{\rm ejecta}$
    ($10^{50}\,{\rm erg})$ \\
\hline
1.4+1.4 & 0.022 & 1.0 \\
1.6+1.2 & 0.043 & 3.6 \\
1.8+1.2 & 0.041 & 2.0 \\
\hline
\end{tabular}
\label{tab:nsmerger}
\end{table}

We assume the energy deposition from the decay of radioactive material
given by \cite{metzger10}, which decays with time ($t^{-\alpha}$ with
$\alpha \approx 1.3$).  As in many of the light-curve calculations of
Type~Ia supernovae, we assume this energy deposition dominates the
temperature structure of the ejecta (with homologous outflows into the
low-density media expected in neutron star mergers, shocks are not 
important).

 The temperature profile is calculated using the energy from the
  radioactive decay.  We include a simplified 1-temperature diffusion
  approximation to incorporate the radiation energy transport within
  our coarse transport grid.  This 1-T diffusion scheme has been
  compared to analytic prescriptions for light curves of Type~Ia
  supernovae~\citep{arnett16} and pair-instability supernovae (de la
  Rosa 2017, in preparation).
  With the temperature profiles set by this energy, we
  calculate the macronova spectra and luminosities via ray-trace
  post-process on the opacities.  We linearly interpolate our opacity
database over density and temperature to estimate the opacity for each
transport zone.  For the coarse resolution in our 2-dimensional
simulations, the velocity change across a zone is $\Delta v/c
\sim$~0.005--0.01, which corresponds to a comparable frequency
resolution due to the homologous nature of the flow.
For each zone, we shift the energy of our radiation for each
ray-trace photon energy to the rest frame at zone center of the
ejecta and obtain an opacity from our tables.  As the photons
traverse the zone, they are alternatively bluer/redder than this
zone center.  This method would allow us to reproduce P-Cygni profiles,
but because there are many strong lines in any energy bin, such profiles are
not resolved. As such, the
opacity tables described in Section~\ref{sub:opac_tab} were averaged
over frequency such that $\Delta \nu/\nu$ was around 1\%.  This
procedure resulted in a number of frequency groups ranging from
150--1500, logarithmically spaced between 40--10$^7$~\AA, depending
the temperature.  Averaging within these larger groups was
accomplished by taking an energy-weighted sum of the monochromatic
opacities that are present in the more spectrally resolved tables.
This approach allows us to conduct a wide parameter study while
including a fairly detailed group structure for the atomic physics.
For simplicity, we have assumed that the ejecta is cylindrically
symmetric, comparing the luminosities in only two directions, on axis
and in the radial direction.

   Note that this light-curve scheme is approximate, and these
    results are focused more on providing qualitative features of
    these new opacities.  In an upcoming paper (Wollaeger et al., in
    preparation), we present detailed multi-dimensional, time-dependent
    Monte Carlo calculations of the macronova emission.

\subsection{Comparing Spectra}
\label{subsec:comp_spec}

As indicated above, the approximations made in the atomic physics and
how we implement them can drastically alter the opacity, i.e., the
Sobolev approximation produces an opacity that, over a wide range of
wavelengths, is more than an order of magnitude lower than that predicted
by an approach that smears each line while preserving the overall
opacity. With our transport
simulations, we can calculate how these opacities, and their
implementation, affect the emission predictions from macronovae.

\begin{figure*}
\plottwo{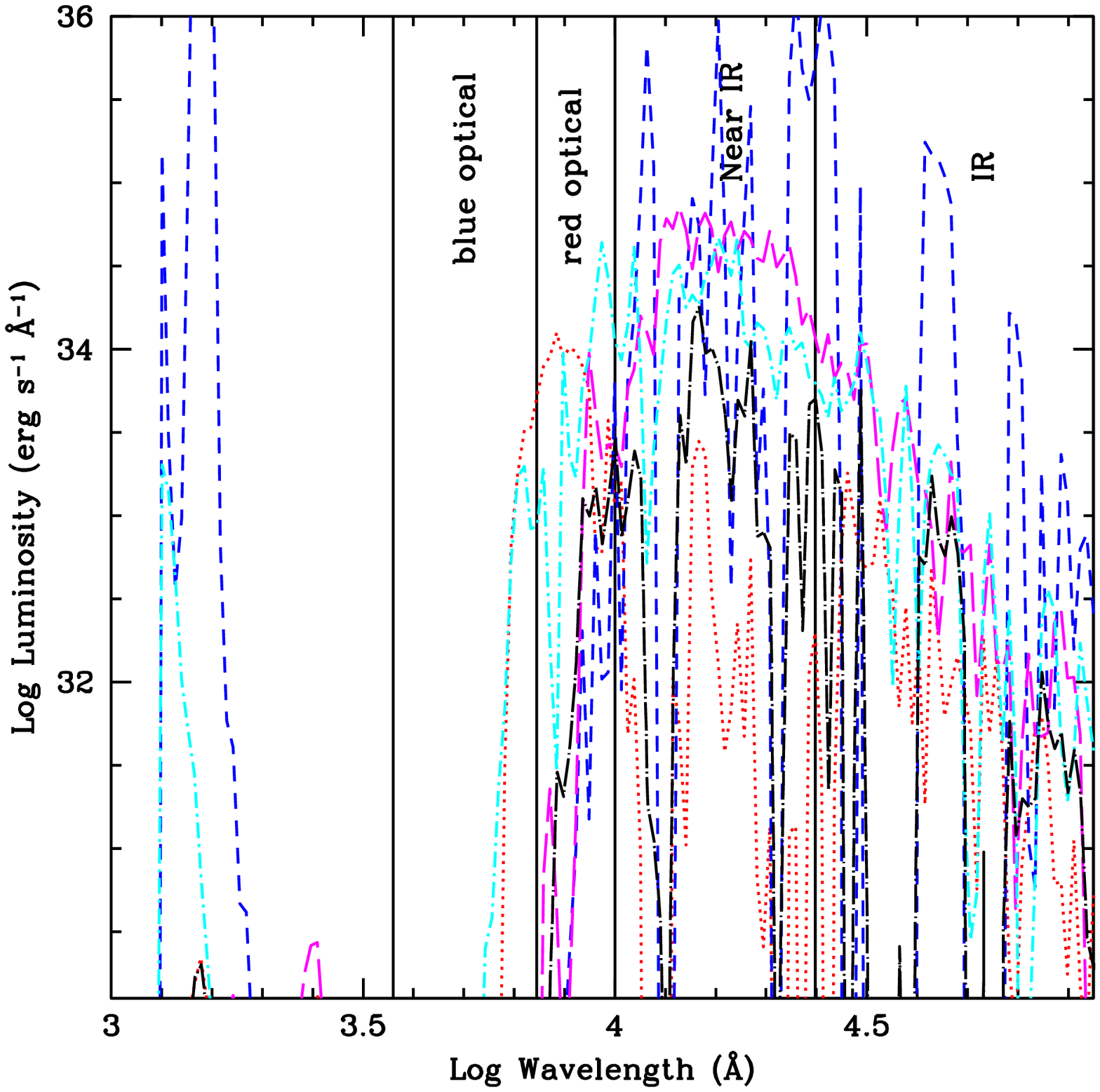}{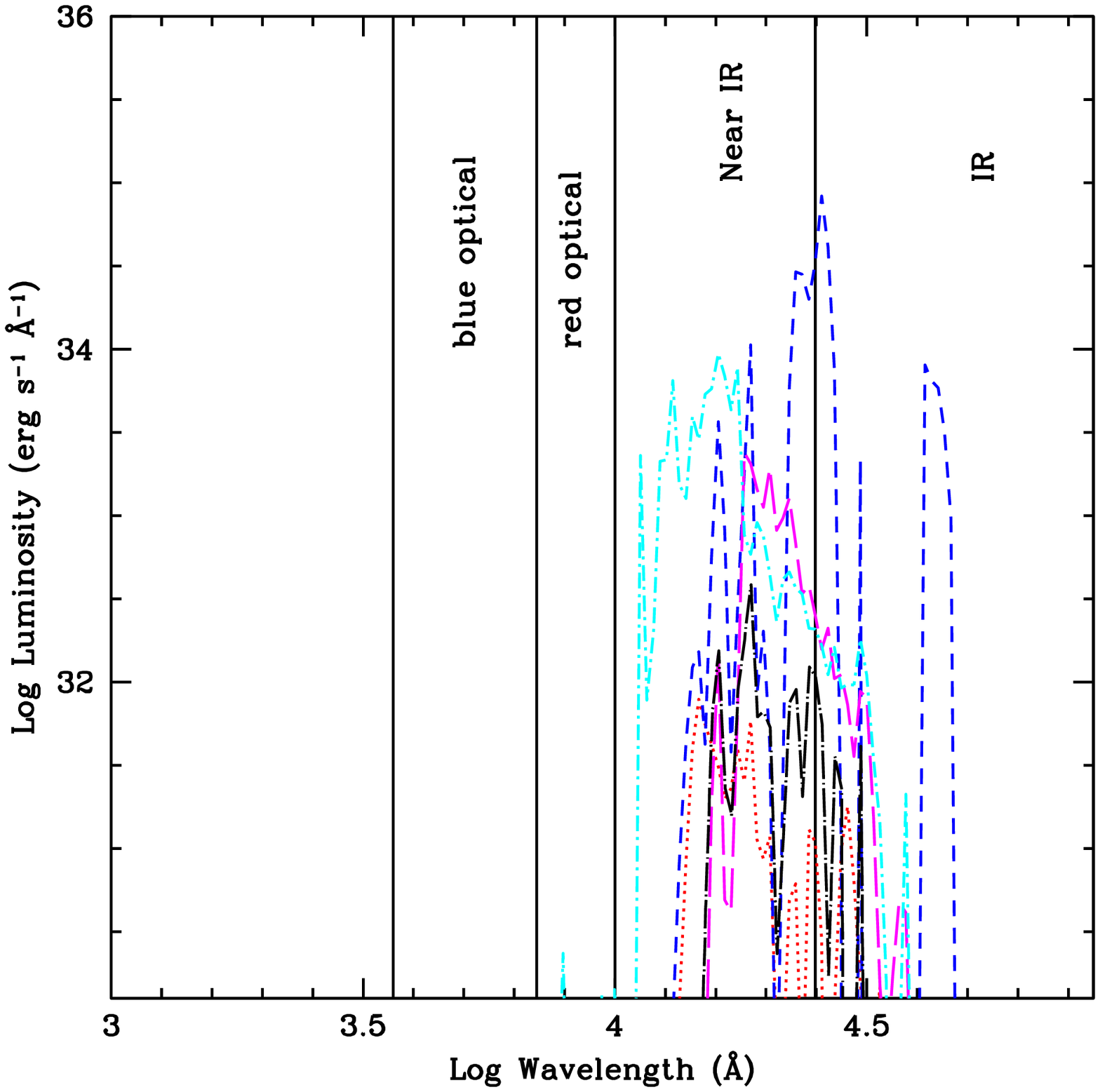}
\caption{Luminosity versus wavelength, assuming single-element species,
  for a merger system consisting of two 1.4\,M$_\odot$ neutron stars at
  1\,d (a) and 10\,d (b).  For these studies, we consider Nd (red), Ce
  (blue), Sm (cyan) and U (magenta) using the semi-relativistic opacities.
  The black curve is produced by
  combining these elements to form a single opacity
  (see text for details).  Gaps in
  the opacity allow higher luminosities at some wavelengths, but
  combining them fills in these windows, reducing the
  luminosities in these wavelength bands.
  }
\label{fig:specel}
\end{figure*}

First, we note that, even though a
minimal amount of smearing will merge many lines in our opacity
calculations, any individual element will have gaps in its opacity.
Figure~\ref{fig:specel} shows the spectra produced if we assume the
opacity is dominated by a single element, with four curves representing Nd,
Ce, Sm, and U.
(The significant UV emission that occurs for Ce between $\sim$~7--10~\AA\ 
is an artifact resulting from the gap between the bound-bound lines and
bound-free edge mentioned previously in Section~\ref{sec:opac}.)
For any individual element, gaps exist that allow flux out of the merged
system. However, when a mixed material, obtained by combining a broad
range of elements, is considered, such gaps will be
filled and the luminosity will be pushed toward longer wavelengths.
As an illustration of this effect, see the black curves
in Figure~\ref{fig:specel}, which represent a mixed material
with combined opacity consisting of 30\% each of Nd, Sm, and Ce, plus 10\% U,
representative of the third $r$-process peak.
In most NSM models, roughly half of the material consists of lighter
elements with much lower opacities, especially in the near infra-red (nIR).
By assuming the stated composition, we overestimate the effect of
lanthanides.  We have modeled an additional test case in which
the opacity was halved,
assuming the lighter elements make up half of the atoms and do not
contribute to the opacity. In this model, the optical contribution can be 2
orders of magnitude brighter in the first 8~hours (although the luminosity
is still dominated by nIR and IR), but it drops significantly after that,
again not contributing to the light curve. The nIR contribution is brighter in
this scenario, but still an order of magnitude dimmer than the IR at
peak. This behavior is studied in more detail in Wollaeger et al.
(in preparation).

At 10\,d, the luminosity is miniscule at short
wavelengths, rising only for photons with wavelengths above 10,000~\AA.
Whether the luminosity exceeds some nominally measurable value
(say, $10^{32}$\,erg\,s$^{-1}$\,\AA$^{-1}$ at 5\,d) above a wavelength
of 11,000~\AA\ or 15,000~\AA\ depends upon which element we choose.  At 5\,d,
the luminosity is above this value for the combined semi-relativistic opacity
only for photons above 15,000~\AA.  These differences define the exact bands
in the infra-red that dominate the opacities.

\begin{figure}
\plotone{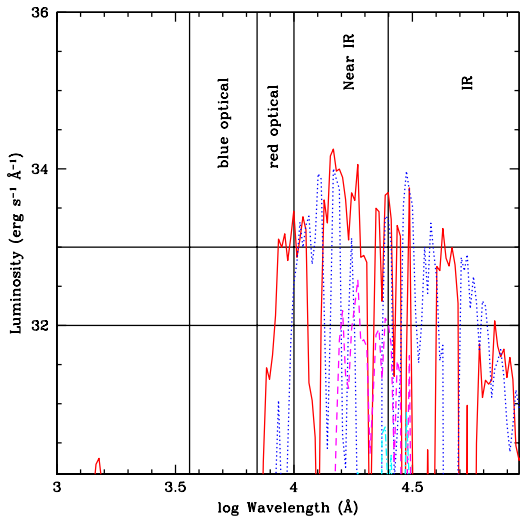}
\caption{Luminosity versus wavelength for our $1.4+1.4\,{\rm M_\odot}$
  merger model at 1 and 10\,d for the semi-relativistic (solid/dashed)
  and the fully relativistic (dotted/dot-dashed) opacities.  The red
  (solid) and magenta (dashed) curves represent the semi-relativistic
  model at 1 and 10\,d, respectively.  The blue (dotted) and cyan
  (dot-dashed) curves represent the fully relativistic model at 1 and
  10\,d, respectively.  The observed spectrum corresponds to what lies
  above some detection limit.  The horizontal lines depict two different
  detection limits ($10^{32},10^{33}$\,erg\,s$^{-1}$\,\AA$^{-1}$),
  respectively.}
\label{fig:speccomp}
\end{figure}

We can use these combined opacities to study the importance of our different
atomic physics models. For example, Figure~\ref{fig:speccomp} compares the
spectra obtained from the combined opacities for the semi-relativistic and
fully relativistic models.  At 2\,d, the luminosity exceeds
$10^{32}$\,erg\,s$^{-1}$\,\AA$^{-1}$ for wavelengths above 13,000~\AA\ 
for the semi-relativistic opacity, compared to 12,000~\AA\ for the fully
relativistic opacity.
This difference in wavelength is, by far, the largest effect on
the spectra due to the various models considered here.
In contrast, the use of opacities from the most-approximate FR-SCR model
produces shifts of less than a few thousand~\AA\ compared to the FR model.

Aside from the shift toward mid-infra-red bands, there are no clear 
spectral lines from these lanthanide opacities, and it may be difficult 
to identify macronovae from their spectra.

\subsection{Light Curves}

These shifts in the light curves produce trends in the broad-band
luminosities.  For these light curves, we consider the luminosities in
four broad bands: the infra-red (we define ``IR'' as photons with
energies between 0.124--0.496\,eV or 10.0--2.5\,$\mu$m), near
infra-red (``nIR" is defined as photons between 0.496--1.24\,eV or
2.5--1\,$\mu$m), red-band optical (defined as photons between
1.24--1.77\,eV or 1--0.7\,$\mu$m), and blue-band optical (defined as
photons between 1.77--3.44\,eV or 0.7--0.36\,$\mu$m).

\begin{figure}
\plotone{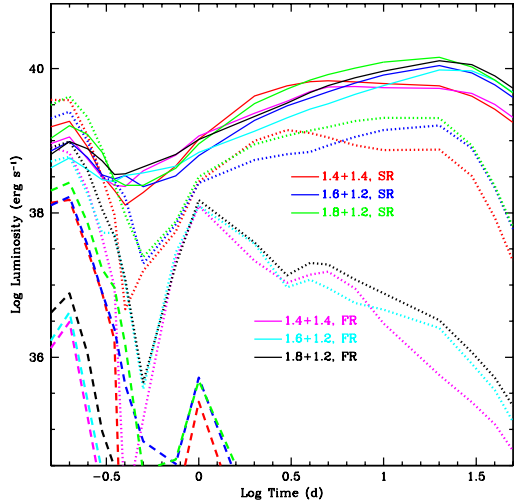}
\caption{IR (solid), nIR (dotted) and red-band optical (dashed)
  luminosities versus time for our three merger models using both the fully
  relativistic and semi-relativistic opacities.  We do not show the
  blue-band optical as it appears only at early times when the ejecta first
  becomes optically thin (the equivalent of shock ``breakout'',
  although this is closer to shock emergence).  This emergence
  emission will be sensitive to the details of the transport
  calculations and we defer this calculation to a more focused study
  on that particular physics.}
\label{fig:lcmer}
\end{figure}

Figure~\ref{fig:lcmer} shows the IR, nIR and red-band optical light curves
for our three
models using both our relativistic and semi-relativistic
approximations.  For our models, the optical band is only bright when
the radiation first emerges, limited to a very short-duration, low
energy burst of photons.  Such results are sensitive to the transport
calculation and we defer their discussion to a later study.  Most of
the radiation is emitted in the infra-red energy range, with a lower
flux in the nIR.  Our ejecta masses vary by a factor of two, and the
resultant luminosities vary by roughly a factor of 3.  Without a
broader range of ejecta masses, we can not firmly determine the
importance of the ejecta mass, but we are consistent with the roughly
linear dependence on this mass determined by \cite{barnes13}.

The blue-band optical emission is extremely low, never peaking above
$10^{35}\,{\rm erg\, s^{-1}}$ and dropping below $10^{30}\,{\rm erg\,
  s^{-1}}$ within a fraction of a day.  The red-band optical is
brighter, peaking above $10^{38}\,{\rm erg\, s^{-1}}$, but also dropping
within a fraction of a day.  The nIR band is most sensitive to the
opacities, where the fully relativistic opacities cause the emission
to decrease sharply after the first day, falling below $10^{37}\, {\rm
  erg\, s^{-1}}$ after about 8\,d.  This behavior is a consequence of
a slight shift in the FR versus SR spectrum; compare the red (solid)
and blue (dotted) curves in Figure~\ref{fig:speccomp} at x-axis values
of 3.7--3.8~\AA.  With the semi-relativistic opacities, this light
curve remains bright (above $10^{39}\, {\rm erg\, s^{-1}}$ for 10\,d).
The IR light curve is much broader and peaks above $10^{40}\, {\rm erg\,
  s^{-1}}$.  Understanding these opacities is critical in determining
the detectors needed to observe these mergers.

The bulk of our luminosity is emitted in the IR and we can compare this luminosity 
to the bolometric light curves from past work. Figure 2 in \cite{barnes13} shows 
the bolometric light curves for a range of simplified models with ejecta masses 
ranging from 0.001--0.1\,M$_\odot$.  In that work, the characteristic 
velocity is parametrized as a fraction of the speed of light, ranging from 0.1--0.3 times the speed 
of light.  Our models are in the massive range of their suite (0.02--0.043\,M$_\odot$) 
with characteristic velocities below 0.1 times the speed of light.  As such, we 
expect our light curves to be broader than most of \cite{barnes13}.  In addition, 
as we have discussed, our opacities are more than an order of magnitude higher 
than those of \cite{barnes13}, further broadening our light curves and producing 
peak fluxes that are roughly an order of magnitude lower.
This order of magnitude decrease in the flux demonstrates the sensitivity
of the predicted light curves to the manner in which the opacities are
calculated.

\begin{figure}
\plotone{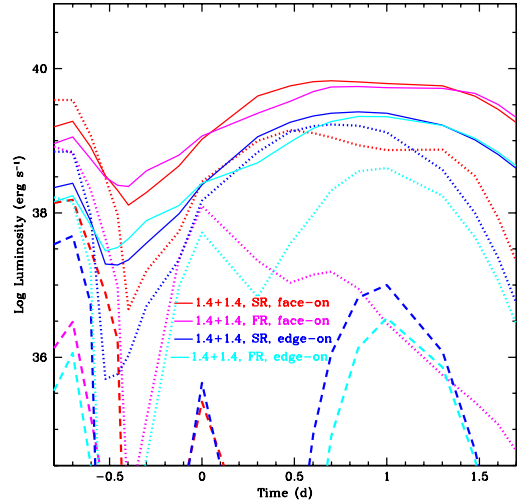}
\caption{IR (solid), nIR (dotted) and red-band optical (dashed) luminosities
  versus time for our $1.4+1.4\,{\rm M_\odot}$ merger using both the
  fully relativistic and semi-relativistic opacities.
}
\label{fig:angle}
\end{figure}

Figure~\ref{fig:angle} shows the variation in the macronova light
curves, comparing the face- and edge-on luminosities of a binary 
NS system composed of two 1.4\,M$_\odot$ components.  Typically, the 
edge on luminosities are 3--4 times lower.  The sensitivity of the 
light curves to shifts in the spectra do allow larger differences 
in the nIR, where the nIR emission from an edge-on burst is over an 
order of magnitude brighter than that of its semi-relativistic counterpart.

\subsection{Caveats}
\label{sub:caveats}

With our SR opacities more than 10--100 times greater in the optical and near-IR
than any current calculation, it is not surprising that our emission
is dimmer than past work.  The comparisons are even more extreme for our
FR opacities, but bear in mind that these light-curve
calculations are designed to demonstrate the variations caused by
the uncertainties in the opacities, and may not be the final macronova
light curves.  A number of uncertainties in the light-curve models 
may yet change the final theoretical light curves.

\begin{figure}
\plotone{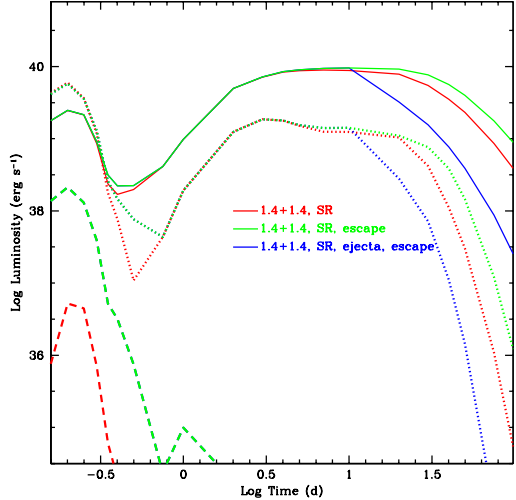}
\caption{Luminosity versus time for three variations on the 1.4+1.4\,M$_\odot$
models.  Comparisons for IR (solid), nIR (dotted) and red-band optical (dashed)
are made between our standard model (red) and two variations that explore the
impact of uncertainties discussed in Section~\ref{sub:caveats}.
These include a model
where gamma-ray deposition is reduced by a factor that varies exponentially in
time after 10\,d (green),
and a model that includes this reduced-gamma deposition, as well
as an increase in the mass of the tidal ejecta (blue).  The increased mass
comes from including material in the tidal ejecta if it has a velocity
greater than one half of the escape velocity, rather than just the escape
velocity.}
\label{fig:lclong.ps}
\end{figure}

Although our ray-tracing transport is not as sensitive to the low
spatial resolution of typical multi-dimensional light-curve
calculations, our simplified transport scheme does not follow the true
temperature evolution of the ejecta.  As the material becomes more
transparent to gamma-rays and, ultimately, electrons emitted during
the decay of our $r$-process elements, the energy deposition will
decrease~\citep{hotokezaka16,barnes16}.  We do not include this effect
and our light curves are almost certainly broader than that produced
by full transport calculations.  As we understand these uncertainties
more fully, we expect shifts in these light curves.  To test the
importance of this effect at late times, we compare our standard model
with one that assumes the energy deposition decreases
  dramatically after 10\,d assuming a larger fraction of the
gamma-rays and energetic electrons escape (see
Figure~\ref{fig:lclong.ps}).  Although artificial, this figure
provides an idea of the importance of accurate decay models.

\begin{figure}
\plotone{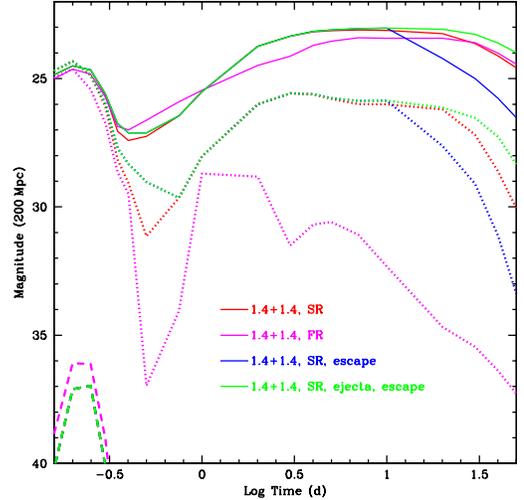}
\caption{Magnitude versus time for the three models shown
in Figure~\ref{fig:lclong.ps}: standard model (red), reduced-gamma deposition
model (blue), and the model with both reduced-gamma deposition and increased
tidal ejecta mass (green).  Additionally, a standard model with the fully
relativistic opacity treatment is included (magenta). Emission from
different wavelength bands are indicated by curves of the following types:
IR (solid), nIR (dotted) and red-band optical (dashed).}
\label{fig:lcmag.ps}
\end{figure}

Additional uncertainties include our initial conditions.  For example,
the amount of ejecta depends both on the exact simulation and
on the time at which we use these simulation
results. In our models, we used the ``ejecta'' produced in the first
18--25\,ms of a merger simulation.  Had we used later-time models,
some of the ``bound'' material could gain energy to become unbound.
To test this hypothesis,
we performed a simulation where the tidal ejecta includes
any material with velocities exceeding half the escape velocity (instead of our
standard full escape velocity).  We assumed that this material does not
decelerate beyond this ejection velocity and recalculated the light
curve.  Figure~\ref{fig:lclong.ps} shows the revised light curve from
these increased ejecta runs.  The corresponding magnitude plot
(assuming the AB magnitude in a band is set to $-2.5
\log_{10}({\rm Luminosity}/(4 \pi d^2)) - 48.6$ where $d=200\,{\rm Mpc}$)
is shown in Figure~\ref{fig:lcmag.ps}. AB magnitudes, instead
of ${\rm erg\, s^{-1}}$, were chosen for this figure as a convenience
to observers. 

Finally, we have assumed that the ejecta are dominated by the $r$-process
ejecta in the tidal tails.  Winds and shocked ejecta from the
newly-formed proto-neutron star (as well as the torus of high
angular-momentum material around this core) can produce more elements
near the iron peak, and these elements have lower opacities,
particularly in the nIR energy bands.  For cases where considerable 
mass is ejected through winds and shocks, the light curves could 
be much brighter.

\section{Summary}

The ejecta from the merger of two neutron stars have the potential to
produce an EM signal that can be used, in conjunction with
a GW detection, to both pinpoint the GW source and better understand
the physics of these mergers.  However, past calculations of the
emission from these ejecta used simplifications to the opacities,
e.g. the expanding-opacity approximation based on the Sobolev method, that
may not be valid for a macronova transient.  Compared to the
Doppler-broadened opacity method proposed here, expansion opacities
are over an order of magnitude smaller.  All of the
atomic physics models considered in this work produce macronova luminosities
that are shifted redward to infra-red bands and an order of magnitude dimmer
compared to past calculations. Discrepancies appear
in the details of the spectra arising from these models, such
as the precise wavelength band in which an observable EM signal occurs,
indicating some sensitivity to the atomic physics assumptions
that should be considered in future studies.
Simplifications associated with the morphology of the dynamic ejecta
and radiation transport scheme applied in this work also
warrant additional investigation.

These results have immediate repercussions for any follow-up searches
for GW sources, arguing that infrared detectors are key. The primary
caveat to this conclusion is that our study has focused exclusively on
the dynamic ejecta of a neutron star merger. Being extremely
neutron-rich and ejected at large velocities, these ejecta produce the
``strong" $r$-process material ($A>130$) for which we have calculated
representative opacities in this paper. Matter that is exposed longer
to the neutrino emission from the remnant, either in the form of
neutrino-driven winds or accretion disk material that becomes unbound
on a viscous time scale ($\sim 0.3$~s), will, however, possess
different properties: it will have lower velocities and larger
electron fractions and therefore undergo a different nucleosynthesis.
A number of recent studies \citep{wanajo14,perego14,just15} found that
compact binary mergers also eject a substantial amount of such matter
with higher electron fractions. This material complements the dynamic
ejecta with lower-mass $r$-process ($A<130$) and produces transients
that are brighter and peak at shorter wavelengths~\citep{kasen15,martin15}.
One also has to seriously consider the
possibility that the purported observations of macronovae, e.g. GRB
130603B~\citep{tanvir13,berger13} and GRB 060614~\citep{jin15,yang15}
had a different origin. Such solutions, however, have their own
difficulties~\citep{berger13} and we leave the further exploration of
the topic to future studies.  Certainly the upper limits placed by GRB
150101B suggest that either the observations from GRB 130603B and GRB
060614 did not arise from macronovae or there is sufficient
variability in the disk wind to produce a wide range of
results~\citep{fong16}.

This work was carried out under the auspices of the National Nuclear Security
Administration of the US Department of Energy at Los Alamos National Laboratory
and supported by contract no DE-AC52-06NA25396.
Helpful discussions with O.~Korobkin are gratefully acknowledged.

\bibliography{master}

\end{document}